\documentclass[%
reprint,
frontmatterverbose, 
nofootinbib,
amsmath,amssymb,
aps,showkeys,showpacs,
prd,
prstab,
prstper,
]{revtex4-1}
\usepackage{newtxmath,newtxtext}
\usepackage{graphicx}
\usepackage[table]{xcolor}
\usepackage[caption=false]{subfig}
\usepackage{dcolumn}
\usepackage{bm}
\usepackage{dsfont}
\usepackage{amsmath,amssymb,stackengine}
\usepackage{siunitx}
\usepackage[caption=false]{subfig}
\usepackage{braket}
\usepackage{listings}
\usepackage{xcolor}
\usepackage{hyperref}
\usepackage{tcolorbox}
\usepackage{braket}
\hypersetup{colorlinks=true,linkcolor=blue,urlcolor=blue,citecolor=blue}
\usepackage{graphicx}
\usepackage{dcolumn}
\usepackage{bm}

\usepackage{orcidlink}
\usepackage{multirow, makecell}
\usepackage{rotating}
\newcolumntype{C}[1]{>{\centering\arraybackslash}p{#1}}

\begin{document}
\title{What solves the Hubble tension in phenomenological dark energy models at background level?}

\author{Manosh T. Manoharan\orcidlink{0000-0001-8455-6951}}
\email{tm.manosh@gmail.com; tm.manosh@cusat.ac.in}
\affiliation{Department of Physics, Cochin University of Science and Technology, Kochi -- 682022, India.}%

\date{\today}

\begin{abstract}
	Few phenomenological models tend to favour higher values of the Hubble parameter, often at the expense of invoking phantom transitions. These models achieve this without introducing additional parameters, akin to the simplicity of the concordance $\Lambda$CDM model. In this work, we investigate two such models -- Phenomenologically Emergent Dark Energy (PEDE) and Granda-Oliveros Holographic Dark Energy (GOHDE) -- to assess how correlations between $H_0$ and $\Omega_m$, as well as the choice of datasets, influence conclusions regarding their potential to address the Hubble tension at the background level. We find that minimally extended versions of these models favour notably low values for the Hubble parameter, with the perceived preference for higher values driven by the associated prior. Excluding BAO Ly$\alpha$-$H(z)$ data points at a redshift of $\sim 2.3$ results in a Hubble parameter that remains in significant tension with SH0ES measurements. In contrast, including these data points favours a higher $H_0$, as they suggest a relatively lower matter density within the framework of the assumed fiducial cosmology. Additionally, recent DESI DR1 and DR2 data exhibit mild tension with BAO-$H(z)$ estimates from SDSS. We demonstrate that the inclusion of stringent constraints, such as the CMB shift-parameter along with Pantheon$^+$, on the effective pressure less matter density significantly impacts the estimation of the Hubble parameter. Finally, reinterpreting these models in terms of interacting dark sectors with $Q = 3H\gamma_{\Lambda}\tilde{\rho}_{m}$ reveals that addressing the Hubble tension necessitates a varying $\gamma_{\Lambda}$ characterised by a singular sign-switch behaviour. This phantom behaviour, or equivalently, the onset of violation of the null energy condition in the future, is crucial for minimal models to solve the Hubble tension. 
\end{abstract}

\maketitle

\section{Introduction}
The most intricate puzzle in cosmology could arguably be the Hubble tension \cite{di2025cosmoverse}, which refers to the discrepancy between the current value of the Hubble parameter reported by local observations, such as \cite{Riess_2022} ($H_0=73.04 \pm 1.04$ km$/$s$/$Mpc), and that inferred from CMB observations ($H_0=67.4\pm0.5$ km$/$s$/$Mpc) \cite{aghanim2020planck}. While local observations depend on calibrators such as Cepheid variables, the relatively high value of $H_0$ reported might stem from calibration issues, as discussed by \cite{freedman2025statusreportchicagocarnegiehubble}. Conversely, CMB observations are grounded in the $\Lambda$ cold dark matter ($\Lambda$CDM) cosmological model, raising questions about the validity of $\Lambda$ as well as CDM. In either case, until more statistically robust data becomes available, the claims in \cite{freedman2025statusreportchicagocarnegiehubble} may remain speculative, as detailed by \cite{Li_2024}. Furthermore, as compelling evidence against $\Lambda$ continues to emerge, as suggested by \cite{desicollaboration2025desidr2resultsii}, the reliability of CMB-based observations could also be challenged. Resolving the Hubble tension, therefore, is far from straightforward. While the observational community focuses on scrutinising calibration methods, any potential flaws in \cite{Riess_2022} could necessitate a downward revision of the Hubble parameter. On the other hand, the theoretical community remains invested in exploring alternatives to the $\Lambda$CDM model and investigating the possibility of ‘new physics’ to account for a higher Hubble parameter. Within this domain, dark energy models play a pivotal role in addressing the Hubble tension, albeit often at the cost of introducing exotic features such as phantom crossings. Recent DESI results, indicating an evolving dark energy component with a phantom crossing \cite{desicollaboration2025desidr2resultsii, scherer2025challenging}, lend further credibility to such models.

Among the extensive array of dark energy models, only a select few are inherently capable of predicting a higher value of the Hubble parameter at the background level, such as the phenomenological emergent dark energy (PEDE) model proposed by \cite{Li_2019}. As might be intuitively anticipated, the PEDE model resolves the Hubble tension at the background level through a dynamic dark energy equation of state parameter that exhibits phantom behaviour. However, the reliance on phantom behaviour to address the Hubble tension represents arguably the most significant (and perhaps the sole) challenge in employing dark energy as a solution at the background level. Here, we will not attempt to resolve this issue; instead, we will focus on identifying the specific features that explicitly enable these models to address the Hubble tension.

If a model can resolve the Hubble tension at the background level—solely by considering the expansion history—it must predict a higher value of the Hubble parameter using datasets that provide the Hubble flow $H(z)$ at various redshifts ($z$). This is precisely what the PEDE model proposed by \cite{Li_2019} offers. However, upon closer inspection, it becomes evident that the results heavily depend on the specific datasets employed. For instance, a quick parameter estimation using only the BAO Galaxy clustering $H(z)$ data for redshifts below 2 yields $H_0=66.94^{+2.87}{-2.93}$ and $\Omega_m=0.38^{+0.07}{-0.06}$ for the PEDE model. These results appear to show tension in both $H_0$ and $\Omega_m$ estimations.This raises the pertinent question: \textit{What resolves the Hubble tension in phenomenological dark energy models at the background level?} To address this in a broader context, we examine both the PEDE model and a comparable framework, the Granda-Oliveros Holographic Dark Energy (GOHDE) model \cite{GRANDA2008275}, which can address the Hubble tension without introducing an additional free parameter, as required by PEDE. Our analysis delves into the finer aspects of these models with minimal extensions to determine what fundamentally resolves the Hubble tension. To this end, we employ various datasets that are commonly used to study the expansion history of the Universe at the background level. Interestingly, despite their different theoretical origins, the impact of the data on the analysis for both models appears remarkably similar, if not identical.

The structure of this manuscript is as follows. First, we introduce the dark energy models of interest -- the PEDE and GOHDE models -- as minimal extensions of the $\Lambda$CDM framework. Subsequently, we explore various re-parameterizations of these models, as proposed by \cite{VONMARTTENS2020100490}. Next, we analyse the influence of different datasets on the models' performance in addressing the Hubble tension, highlighting the critical role played by the correlation between free parameters and the assumed fiducial cosmology in the extraction of certain data points, shaping the derived conclusions at the background level. We then examine the specific features of these minimal extensions. Our findings reveal that these extensions do not favour a significantly large value of $H_0$ in the full picture. By exploring the models within the interacting dark sector framework, we demonstrate that a future violation of the null energy condition is sufficient to resolve the Hubble tension. This conclusion remains consistent across various extensions and manifests as a pole in the evolving interaction strength.

\section{Minimal extensions to $\Lambda$CDM}

Even before the discovery of accelerated expansion, the concordance cosmological constant ($\Lambda$) faced numerous theoretical challenges \cite{RevModPhys.61.1}. Following the confirmation of the Universe's recent accelerated expansion, a plethora of dynamic dark energy models emerged in the literature (see the review by \cite{RevModPhys.75.559} for an overview, and the updated discussion on the dark energy equation of state by \cite{Escamilla_2024}), continuing to this day. Although the original motivation for introducing these alternatives to $\Lambda$ was to address the cosmological constant problem, the advent of cosmological tensions provided them with an alternative purpose.
One of the key challenges in replacing $\Lambda$ with a dynamic dark energy model has been the stringent observational constraints on the dark energy equation of state parameter ($w_{\Lambda}$). Despite revealing various tensions, most datasets have consistently preferred a value for $w_{\Lambda}$ close to $-1$, indicating a strong preference for $\Lambda$. However, recent results from DESI \cite{karim2025desi} suggest a $\sim4.2\sigma$ preference for a varying dark energy component over $\Lambda$. Perhaps the most intriguing aspect of the findings by \cite{karim2025desi} is the suggestion of a phantom nature for dark energy. To reconcile the matter density ($\Omega_m$) with an increased Hubble parameter, additional degrees of freedom in the expansion history become necessary. This introduces variants like $w$CDM models, where $w=w_\Lambda<-1$ can push the Hubble parameter to larger values while maintaining a relatively unchanged matter density. At the background level, this necessitates dynamic dark energy exhibiting some phantom behaviour, rooted in the correlations between $H_0$, $\Omega_m$, and $w_\Lambda$. Interestingly, if we were free to lower $\Omega_m$, even the $\Lambda$CDM model could produce a higher value for $H_0$. However, this approach is undesirable as such an abrupt change in $\Omega_m$ would distort the acoustic peak positions in the CMB power spectrum, which currently align well with the concordance model's predictions.

In the simplest $w$CDM models, the dark energy equation of state is typically treated as a free parameter, with no a priori reason to fix it to a particular value, such as $w_\Lambda = -1.314$. However, setting $w_\Lambda = -1 - (\pi/10)$, at least from a phenomenological standpoint, presents intriguing possibilities. In this context, we shall consider two phenomenological dark energy models which, upon fixing a few free parameters, reduce to minimal $\Lambda$CDM-like models without introducing any additional degrees of freedom. See \cite{PhysRevD.102.023518} other interesting minimal extensions.

\subsection{The Phenomenological Emergent Dark Energy (PEDE)}

Reference \cite{Li_2019} introduced a novel phenomenological form of dark energy, which they term \textit{emergent}. Unlike matter or radiation, which dilute as the Universe expands, or the cosmological constant ($\Lambda$), which remains constant throughout, this dark energy density was negligible in the past and has only recently emerged. Consequently, the phenomenological emergent dark energy (hereafter PEDE) offers a distinctive approach to addressing the Hubble tension without significantly altering early-Universe physics. Assuming the flat Friedmann -- Lema\^itre -- Robertson -- Walker (FLRW) metric, they arrived at the Hubble flow as,
\begin{equation}
H^2(z)=H_0^2\left[\Omega_m(1+z)^3+\Omega_\Lambda(z)\right],
\end{equation}
where
\begin{equation}
\Omega_\Lambda(z)=\left(1-\Omega_m\right)\lbrace 1-\tanh\left[\log_{10}\left(1+z\right)\right]\rbrace.
\end{equation}
This construction allows one to reconstruct the dark energy equation of state parameter as,
\begin{align}
w_\Lambda(z)&=-1+\frac{(1+z)}{3}\frac{d}{dz}\left[\ln\Omega_\Lambda(z)\right]\nonumber\\
&=-1-\frac{1}{3\ln10}\lbrace 1+\tanh\left[\log_{10}\left(1+z\right)\right]\rbrace
\end{align}
This model naturally leads to an asymptotic de Sitter phase with a gradually weakening phantom dark energy component. As highlighted by \cite{Li_2019}, it predicts a higher value for $H_0$ without significantly altering the matter density, making it a promising candidate for resolving the Hubble tension. A key assertion in their work is the absence of additional free parameters. However, the reliance on a base-10 logarithm in the functional form of $\Omega_\Lambda(z)$ raises concerns. Theoretically, the base of the logarithm introduces a free parameter, $\nu = 1 / \ln(\text{base})$, when using natural logarithms elsewhere. This issue was promptly addressed in subsequent works by \cite{Li_2020}, \cite{PhysRevD.104.063521}, and more recently by \cite{nelleri2023observational}. Consequently, the modified expression is,
\begin{equation}
\Omega_\Lambda(z)=\left(1-\Omega_m\right)\lbrace 1-\tanh\left[\nu\ln\left(1+z\right)\right]\rbrace
\end{equation}
Thus, the original PEDE model emerges when $\nu=1/\ln10\sim0.43$, while $\Lambda$CDM becomes a limiting case at $\nu=0$, corresponding to the base approaching zero. Additionally, for $\nu<0$, the base lies between zero and unity. Consequently,
\begin{equation}
\nu=\begin{cases}
=0, & \Lambda\text{CDM, with base}=0,\\
<0,  & 0<\text{base}<1,\\
>0, &\text{base}>1.
\end{cases}
\end{equation}
It is crucial to observe that the PEDE model represents a specific case where the base is $10$. This highlights that the free parameter, akin to the dark energy equation of state in the $w$CDM model, plays a pivotal role in resolving the Hubble tension. Notably, a positive $\nu$ has the potential to alleviate this tension, whereas a negative $\nu$ may exacerbate it. Thus, investigating the PEDE model or its generalised version becomes a pursuit to constrain the value of $\nu$. In our analysis, we shall initially adopt a non-negative Dirac and uniform prior for $\nu$ and subsequently broaden the scope to encompass negative values.

\subsection{Granda--Oliveros Holographic Dark Energy (GOHDE)}

In \cite{GRANDA2008275} authors introduced a new causally consistent infrared cut-off for a distinct class of dark energy models known as Holographic Dark Energy (HDE). For an extensive review of HDE, refer to \cite{WANG20171}. Despite criticisms regarding the foundational principles of HDE, as highlighted by \cite{manoharan2025entropy}, the model has been thoroughly examined and extended.
Among these extensions, the model employing the causally consistent infrared cut-off with the Hubble parameter and its derivative is termed the Granda–Oliveros Holographic Dark Energy (GOHDE) model. This approach incorporates two additional free parameters beyond $\Lambda$CDM, introduced through the infrared cut-off. One of these parameters can be redefined as the current value of the dark energy equation of state, while the other emerges as the tracker parameter, which determines whether the model aligns with the kinematics of other cosmic fluids. The characteristics of this model, along with the potential for phantom crossing accompanied by negative energy density, have been explored in detail by \cite{Manoharan2024}.

The Hubble parameter for the most generalised form of the GOHDE model within a non-flat FLRW metric is expressed as,
\begin{align} H^2(z)= H_0^2&\left(\frac{{\Omega _{r}}e^{-4 x}}{-\alpha +2 \beta +1}+\frac{2 {\Omega _{m}}e^{-3 x}}{-2 \alpha +3 \beta +2}+\frac{ {\Omega _{k}}e^{-2 x}}{-\alpha +\beta +1}\right.\nonumber \\&\left.\quad +\text {C}_1e^{-\frac{2 (\alpha -1) x}{\beta }}\right), 
\end{align}
with 
\begin{align} \text {C}_1&=\frac{\alpha +{\Omega _{k}}}{\alpha -\beta -1}-\frac{2 {\Omega _{m}}}{-2 \alpha +3 \beta +2}+\frac{{\Omega _{r}}}{\alpha -2 \beta -1}\nonumber \\&\quad +\frac{\beta }{-\alpha +\beta +1}+\frac{1}{-\alpha +\beta +1}.
\end{align}
We can redefine $\alpha$ by expressing it in terms of the current value of the dark energy equation of state as,
\begin{align}
w_{z0}=\frac{-2 \alpha (\Omega _k+1)}{3 \beta (\Omega _m+\Omega _r-1)}+\frac{2 \Omega _k+\Omega _r+3}{3 (\Omega _m+\Omega _r-1)}-\frac{2}{3\beta }.\nonumber \\ 
\end{align}
which gives us
\begin{align}
\alpha&=~\frac{\beta \left[ -3 w_{z0} (\Omega _{m}+\Omega _{r}-1)+2 \Omega _{k}+\Omega _{r}+3\right] }{2 (\Omega _{k}+1)}\nonumber \\&\quad -\frac{(\Omega _{m}+\Omega _{r}-1)}{ (\Omega _{k}+1)}.
\end{align}
Given our focus on a model that serves as a minimal extension of $\Lambda$CDM, we can simplify the analysis by assuming a flat metric and neglecting the effects of radiation. This leads us to the expression, 
\begin{align}
H^2(z)=H_0^2\left[\tilde{\Omega}_{m}(1+z)^3+(1-\tilde{\Omega}_{m})(1+z)^{-3(1+w_\Lambda)}\right],
\end{align}
with 
\begin{align}
\tilde{\Omega}_{m}&=\Omega_{m}\left(\frac{2}{2-2\alpha+3\beta}\right)\text{ and }w_\Lambda=-1+\frac{2\left(\alpha-1\right)}{3\beta}.
\end{align}
In essence, the GOHDE model can be viewed as a $w$CDM model with a modified matter component. However, if we treat the matter density as $\Omega_m$ itself, the dark energy exhibits tracking behaviour. In this work, we focus on fixing the present value of the dark energy equation of state, $w_{z0}$, to $-1$, thereby reducing the model to a single free parameter, $\beta$. Notably, when $\beta = \frac{2}{3}$ and $\alpha = 1$, both the matter density and $w_\Lambda$ revert to their $\Lambda$CDM counterparts. More precisely, the GOHDE model reduces to $\Lambda$CDM upon neglecting radiation and curvature and setting $\beta = \frac{2}{3}$. Given the general construction of HDE, this approximation will not significantly impact our analysis. Fixing $w_{z0} = -1$ then allows us to substitute
\begin{align}
\alpha=1+\Omega_m\left(\frac{3\beta}{2}-1\right).
\end{align}
This enables us to re-express the free parameters in the Hubble flow as,
\begin{align}
\tilde{\Omega}_{m}&=\frac{2\Omega_m}{3\beta-3\beta\Omega_m+2\Omega_m}\\
w_\Lambda&=-1+\Omega_m-\frac{2\Omega_m}{3\beta}
\end{align}
Setting $\beta = \frac{2}{3}$ once again recovers the $\Lambda$CDM model, whereas values of $\beta < \frac{2}{3}$ predict a phantom-like future. Earlier, reference \cite{Manoharan2024} treated $\beta$ as a free parameter and demonstrated the possibility of negative dark energy density in the past. Their best-fit values of $\beta$ were found to be very close to $\frac{2}{3}$, allowing dark energy to mildly mimic matter with negative energy density. However, this did not provide an automatic resolution to the Hubble tension.

In the following sections, we will show that choosing $\beta = \frac{1}{3}$ favours a higher value of $H_0$, analogous to the behaviour observed in the PEDE model. Just as the generalised PEDE reduces to the PEDE introduced by \cite{Li_2019}, setting $\beta = \frac{1}{3}$ in GOHDE can effectively elevate the Hubble tension without any additional free parameters as similar to PEDE. A key distinction in the GOHDE model is that, unlike PEDE, $\Omega_m$ no longer represents the total matter density. The dark energy equation of state can approach zero in the past, behaving similarly to matter. Consequently, it is the effective matter density, denoted by $\tilde{\Omega}_m$, that truly accounts for the matter content. Therefore, although we fit for $\Omega_m$, the effective matter density governing dynamics is $\tilde{\Omega}_m$. Recognising this distinction is crucial for data analysis, particularly when using observables that constrain matter density explicitly, such as the CMB shift parameter.

\section{Interaction Picture of dynamical dark sector}
Having introduced the two models of interest, it is essential to examine the degeneracies present within them, not only among their free parameters but also in relation to other well-established dark energy frameworks. For example, the equation above clearly indicates that, at the background level, GOHDE exhibits degeneracy with a modified $w$CDM model. Previous studies, such as \cite{VONMARTTENS2020100490} and the recent follow-up by \cite{TAMAYO2025101901}, have demonstrated that numerous dynamical dark energy models share such degeneracies. We therefore anticipate observing similar degeneracies within the PEDE and GOHDE models. Our goal, however, is to leverage these degeneracies to identify common features in how these models address the Hubble tension. To this end, we will reinterpret our models within the framework of interacting dark sector scenarios, wherein dark energy and dark matter can exchange energy and transform into one another. See \cite{shah2025interacting, 10.1093/mnras/stae2712, Yang_2018, PhysRevD.101.063502, silva2025new} for additional references. 

Following the framework established in \cite{TAMAYO2025101901}, the continuity equations for a non-interacting dynamical dark energy component are given by,
\begin{align}
\dot{\rho}_m+3H\rho_m=0,\text{ and }
\dot{\rho}_{\Lambda}+3H(1+w_{\Lambda})\rho_{\Lambda}=0.
\end{align}
In the interacting scenario, we denote these densities by $\tilde{\rho}_i$ to clearly distinguish them. The corresponding continuity equations with interaction are then expressed as,
\begin{align}
\dot{\tilde{\rho}}_m+3H\tilde{\rho}_m=\tilde{Q},\text{ and }
\dot{\tilde{\rho}}_{\Lambda}=-\tilde{Q}.
\end{align}
To establish the equivalence between these models and the continuity equations above, it is necessary to recognise a relationship between the energy densities and the interaction term $\tilde{Q}$. Consequently, the interaction term will depend on the $\rho_i$ along with the Hubble parameter $H$. Within this framework, we assume that regardless of the specific continuity equations, the total energy density remains unchanged and is given by the sum of the individual components. Under this assumption, the interaction term can be derived as shown in \cite{TAMAYO2025101901}. Thus,
\begin{equation}
\tilde{Q}=3H\tilde{f}\times\left(\frac{\tilde{\rho}_{\Lambda}\tilde{\rho}_{m}}{\tilde{\rho}_{\Lambda}+\tilde{\rho}_{m}}\right),
\end{equation}
where, 
$\tilde{f}=1+[{a\tilde{r}'}/({3\tilde{r}})]\text{ and }\tilde{r}={\tilde{\rho}_m}/{\tilde{\rho}_{\Lambda}}$
The prime symbol denotes differentiation with respect to the scale factor, $a$. Using this notation, one can express the interaction term $\tilde{Q}$ in various meaningful forms and subsequently determine the interaction strength. For example, we may write,
\begin{align}
\tilde{Q}&=3H\tilde{f}\times\left(\frac{\tilde{\rho}_{\Lambda}}{\tilde{\rho}_{\Lambda}+\tilde{\rho}_{m}}\right)\tilde{\rho}_{m}=3H\gamma_{\Lambda}\tilde{\rho}_{m},\\
\tilde{Q}&=3H\tilde{f}\times\left(\frac{\tilde{\rho}_{m}}{\tilde{\rho}_{\Lambda}+\tilde{\rho}_{m}}\right)\tilde{\rho}_{\Lambda}=3H\gamma_{m}\tilde{\rho}_{\Lambda}.
\end{align}
These explicit forms involve an interaction strength, denoted by either $\gamma_{\Lambda}$ or $\gamma_{m}$, both of which could depend on the scale factor. It is straightforward to see that $\gamma_{\Lambda} = \tilde{f}\times\tilde{\rho}_{\Lambda}/\left({\tilde{\rho}_{\Lambda} + \tilde{\rho}_{m}}\right)$, and equivalently, the interaction term can be rewritten as $3H \gamma_{m} \tilde{\rho}_{\Lambda}$, where $\gamma_{m} = \tilde{f}\times\tilde{\rho}_{m}/\left({\tilde{\rho}_{\Lambda} + \tilde{\rho}_{m}}\right)$. In either formulation, the background dynamics remain unchanged. The physical significance of $\tilde{Q}$ lies in its sign: a positive value indicates energy transfer from dark energy to dark matter, while a negative value implies the reverse. This characteristic is key to understanding and identifying a distinctive signature associated with the mechanisms that elevate the Hubble tension in dark energy models.

Following the approach outlined in \cite{TAMAYO2025101901}, the PEDE and GOHDE models can be reformulated as interacting models with specific expressions for the interaction term $\tilde{Q}$. It is important to note that the interaction depends on the energy densities defined within the interacting framework, which can be expressed as functions of the energy densities in the non-interacting picture. This transformation is given by,
\begin{align}
\tilde{\rho}_{\Lambda}&=-w_{\Lambda}(a)\rho_{\Lambda}\text{ and }
\tilde{\rho}_{m}=\rho_{m}+\left[1+w_{\Lambda}(a)\right]\rho_{m}
\end{align}
This result rests on the assumption that the total effective equation of state remains identical, as one would expect for models that are degenerate at the background level. By employing the expressions for the energy densities and the dark energy equation of state corresponding to PEDE and GOHDE, one can derive the form of the interaction term $\tilde{Q}$. As noted earlier, we will consider two formulations for $\tilde{Q}$, leading to the interaction strengths $\gamma_{\Lambda}$ and $\gamma_{m}$ for both PEDE and GOHDE. Since our focus is on redshift rather than the scale factor, the corresponding expressions are given by, for PEDE,
\begin{widetext}
\begin{align}
\gamma_{\Lambda}(z)=&\frac{v (\Omega_m-1) \lbrace 2 v \tanh \left[v \log (z+1)\right]+3\rbrace \text{sech}^2\left[v \log (z+1)\right]}{3 v (\Omega_m-1) \text{sech}^2\left[v \log (z+1)\right]+9 \Omega_m (z+1)^3},\\
\gamma_{m}(z)=&-\frac{2 v+3}{v \tanh \left[v \log (z+1)\right]+v+3}-\frac{2}{3} v \tanh \left[v \log (z+1)\right]+1,
\end{align}
\end{widetext}
In both cases, the parametrizations evolve as functions of redshift, with the second formulation notably independent of the matter density. Similarly, for GOHDE,
\begin{align}
\gamma_{\Lambda}(z)&=\frac{2 \Omega_m-3 \beta  (\Omega_m-1)}{3 \left[\beta -\frac{2 \beta  \left(\frac{1}{z+1}\right)^{\left(3-\frac{2}{\beta }\right) \Omega_m-3}}{(3 \beta -2) (\Omega_m-1)}\right]},\\
\gamma_{m}(z)&=\Omega_m-\frac{2 \Omega_m}{3 \beta }.
\end{align}
Rewriting these models offers little advantage in terms of understanding the background evolution itself, but it proves valuable in identifying the features that resolve the Hubble tension. We will return to this analysis after examining the models using specific data sets.

\section{Data and Analysis Methodology}
When can we assert that a model effectively addresses the Hubble tension? At an elemental level, the criterion is straightforward: the model must predict a present-day value of the Hubble parameter ($H_0$) that closely aligns with local observations derived from Cepheid variables. Specifically, the model should estimate $H_0$ at $\sim73$ km$/$s$/$Mpc, with an uncertainty $\sim1$ km$/$s$/$Mpc.
The objective, therefore, is relatively simple: consider a dataset for which the standard $\Lambda$CDM model predicts a lower $H_0$, typically around the Cosmic Microwave Background (CMB) estimate of $67$ km$/$s$/$Mpc, and determine whether the proposed model yields a higher $H_0$ using the same dataset.
Any claim of resolving the Hubble tension must typically withstand rigorous scrutiny, including a full fit to the CMB power spectrum within the framework of the proposed model. However, such an analysis necessitates perturbation-level studies rather than background-level ones, which are computationally intensive and resource-demanding. Consequently, this work will focus on datasets primarily involving background evolution alone.

It is worth noting that the same model can produce different estimates of its free parameters when applied to varying datasets. Additionally, correlations between free parameters may influence both the analysis and the conclusions drawn. In this section, we will outline the datasets employed and the methods used for our study.

\subsection{Data}
\paragraph{\textbf{Cosmic Chronometers}}\textbf{(CC)}: These represent estimates of the Hubble parameter at various redshifts. A well-known compilation of CC data includes 31 data points, extending up to a redshift of nearly $2$. The estimation of these points is based on the relationship $(1+z)H(z)=-{dz}/{dt}$, as detailed in \cite{Jimenez_2002}. These data points are obtained from \cite{DanielStern_2010, Moresco_2016, Zhang_2014, M.Moresco_2012, 10.1093/mnrasl/slv037, 10.1093/mnras/stx301}.\\

\paragraph{\textbf{BAO Galaxy Clustering}}\textbf{(BAO Gal):} Like the CC dataset, this compilation also provides estimates of the Hubble parameter at various redshifts. The dataset used here includes 22 data points spanning redshifts below $1$. These BAO galaxy clustering measurements are derived from the SDSS-III BOSS DR12 galaxy sample, the WiggleZ Dark Energy Survey, and the SDSS DR7 LRG sample \cite{10.1093/mnras/stx721, 10.1093/mnras/stu523, 10.1111/j.1365-2966.2012.21473.x, 10.1093/mnras/stx1090, 10.1093/mnras/stu111, 10.1093/mnras/stu523, 10.1093/mnras/stx721, 10.1093/mnras/stt1290, 10.1111/j.1365-2966.2009.15405.x}. \\

\paragraph{\textbf{BAO Lyman-$\alpha$}}\textbf{(BAO \boldmath Ly$\alpha$)}: This dataset provides an estimate of the Hubble parameter with a significantly smaller error bar at a redshift close to $2.3$. Technically, these data points are related, as they are all derived from SDSS and BOSS. However, due to differences in protocols and the samples used for analysis, they are treated as distinct data points. Recent DESI DR1 and DR2 results also provide estimates at these redshifts, which will be included in the analysis. Given the importance of these data points in shaping the conclusions on the Hubble tension, we shall explicitly list them here.
\begin{table}[h]
		\renewcommand{\arraystretch}{1.3}
	\begin{ruledtabular}
	\begin{tabular}{ccl}
		$z$&$H(z)$ (km$/$s$/$Mpc)&Source\\
		\hline
		$2.3$&$224\pm8$&\cite{refId0Busca} 2013\\
		$2.36$&$226\pm8$&\cite{Font-Ribera_2014} 2014\\
		$2.34$&$222\pm7$&\cite{refId0Delubac} 2015\\
		$2.33$&$224\pm8$&\cite{refId0Bautista} 2017
	\end{tabular}
\end{ruledtabular}
\caption{BAO Ly$\alpha$ data points for the Hubble parameter near redshift $z \sim 2.3$. These estimates primarily assume a drag radius $r_d$ based on the CMB data available at the time.\label{tab:BAOLyaAllOld}}
\end{table}
The dataset mentioned above pertains to earlier surveys and analyses, whereas more recent compilations provide slightly different estimates. For the data considered here, the average redshift is approximately $z \sim 2.33$, and the Hubble parameters are estimated using a drag radius of $r_d = 147.09 \pm 0.26$ Mpc, as derived from Planck 18 \cite{aghanim2020planck}. In the fiducial cosmology assumed, the matter density $\Omega_m=0.27$ and $H_0=70$ km$/$s$/$Mpc.
\begin{table}[h]
	\renewcommand{\arraystretch}{1.3}
\begin{ruledtabular}
	\begin{tabular}{lcl}
		Survey&$H(z=2.33)$ (km$/$s$/$Mpc)& Source\\
		\hline
		SDSS-IV Ly$\alpha$-QSO&$224.47\pm8.41$&\cite{PhysRevD.103.083533} 2021\\
		SDSS-IV Ly$\alpha$-QSO$^+$&$226.71\pm4.81$&\cite{duMasdesBourboux2020} 2020\\
		DESI DR1 Ly$\alpha$-QSO&$239.22\pm4.79$&\cite{Adame_2025} 2025\\
		DESI DR1 Ly$\alpha$-QSO$^+$&$236.12\pm2.79$&\cite{karim2025desi} 2025
	\end{tabular}
\end{ruledtabular}
\caption{BAO Ly$\alpha$ data points for the Hubble parameter near redshift $z \sim 2.33$. The survey labelled ``Ly$\alpha$-QSO$^+$'' refers to the combined analysis of Ly$\alpha$-QSO and Ly$\alpha$-Ly$\alpha$ correlation data. These measurements assume a drag radius $r_d \approx 147.09$ Mpc, based on Planck 18 CMB (TT, TE, EE + low E + lensing) \cite{akarsu2025dynamical}.\label{tab:BAOLyaDESI}}
\end{table}
All the datasets presented in TABLES (\ref{tab:BAOLyaAllOld}) and (\ref{tab:BAOLyaDESI}) provide direct estimates of H(z)H(z). Consequently, parameter estimation relies solely on the expression for the Hubble parameter corresponding to the model under consideration, as discussed in the preceding sections.\\

\paragraph{\textbf{Pantheon Plus}}\textbf{(Pan$^+$):} Numerous compilations of Type Ia Supernova datasets have proven priceless in probing the universe, beginning with the legacy surveys that first confirmed late-time cosmic acceleration \cite{Riess_1998, Perlmutter1998}. In this work, we focus exclusively on the Pantheon$^+$ compilation, which comprises 1701 data points \cite{Riess_2022}. It is important to note that, without incorporating the SH0ES prior on the local Hubble parameter or the absolute magnitude, supernova data alone cannot effectively constrain $H_0$. Other compilations, such as Union 2.1 \cite{suzuki2012hubble} and the more recent Union 3 \cite{rubin2023union}, employ a Dirac prior on $H_0$, which limits their utility for constraining this parameter. For similar reasons, we do not consider the DES compilation either \cite{Abbott_2024}. Unless one undertakes a full reanalysis starting from the fitting of light curves, publicly available $z$, $\mu$ data may be used only to constrain parameters other than $H_0$.

All SNe Ia datasets comprise the apparent magnitude $\mu$, the corresponding redshift $z$, and the standard deviation in $\mu$ or the covariance matrix. Therefore, our aim is to compute $\mu$ using the relevant cosmological model for the analysis. The apparent magnitude is given by,
\begin{equation}
\mu(z)=5\log_{10}\left[\frac{d_L(z)}{\text{Mpc}}\right]+M+25.
\label{eq:apparentmagnitude}
\end{equation}
Here, $M$ denotes absolute magnitude, requiring calibration using other datasets, and $d_L$ represents luminosity distance given as,
\begin{equation}
d_L(z)=c(1+z)\int_{0}^{z}\frac{1}{H(z')}dz'.
\label{eq:D_L}
\end{equation}
By utilising the Hubble parameter $H(z)$ at a given redshift $z$, we can estimate the luminosity distance $d_L$ and thereby compute the apparent magnitude. Since our focus is on addressing the Hubble tension, we do not incorporate direct measurements of $H_0$ using the SH0ES prior. A notable aspect of earlier studies on the generalised version of PEDE, such as \cite{Li_2020}, is the adoption of a Gaussian SH0ES prior on $H_0$. When this prior was omitted, the generalised PEDE model, relying solely on CMB data, yielded an estimate of $H_0$ lower than that predicted by $\Lambda$CDM. This highlights that both the choice of priors and the datasets employed play a critical role in shaping the conclusions drawn. Redshift dependence of $H_0$ is another way of looking at the tension and is addressed in \cite{Dainotti_2021}.\\

\paragraph{\textbf{CMB Shift Parameter}}\textbf{(CMB):} The use of the CMB shift parameter requires considerable caution. There are various arguments both for and against its application, which is why we provide a more detailed discussion here rather than merely stating its use. The CMB shift parameter is a dimensionless quantity, representing a ratio. It is defined as,
\begin{equation}
R=\frac{{l'}_{1}^{TT}}{l_1^{TT}}
\end{equation}
Here, $l_1^{TT}$ denotes the position of the first acoustic peak in the CMB temperature perturbation spectrum, while the primed quantity corresponds to the same position calculated under the assumption $\Omega_{m} = 1$. This aspect is often regarded as a drawback since it necessitates knowledge of the model both with and without dark energy. However, this can be addressed using Boltzmann codes such as CAMB \cite{CAMBPython}. Assuming the earlier physical processes remain unchanged, we can compute this and employ a more user-friendly form of the shift parameter $R$, given by
\begin{equation}
\bar{R}=\sqrt{\Omega_{m}}\int_{0}^{z_{r}}\frac{H_0}{H(z')}dz'
\end{equation}
Here, $z_r$ denotes the redshift of recombination. Here we use, $z_r=1089.92$ and $bar{R}=1.7502\pm0.0046$ \cite{Chen_2019}. For the time being we are not using the full distance prior given in \cite{Chen_2019}.

A key feature to note is the presence of $\sqrt{\Omega_{m}}$, which fundamentally arises from the original definition based on the cold dark matter (CDM) model without dark energy. In most dark energy models, this expression remains applicable since the background evolution already incorporates dark energy dynamics, with $\Omega_{m}$ retaining the same interpretation as in the $\Lambda$CDM model. However, these changes in the GOHDE model, particularly significant when the other free parameters are fixed. For models such as GOHDE, where dark energy exhibits matter-like behaviour in the past, it is essential to use the effective matter density appearing in the Hubble flow when evaluating the $\sqrt{\Omega_{m}}$ term. Since the shift parameter provides only weak constraints on other free parameters, its primary role is to constrain the matter density. Yet, because the definition involves the model estimate excluding dark energy, the effective matter density must be employed.

In the GOHDE model, which can be regarded as an effective $w$CDM model with modified $\Omega_{m}$ and $w_\Lambda$, it is essential to treat these parameters appropriately. However, an additional caveat arises when an extra free parameter is introduced. If one employs $\Omega_{m}$ instead of the effective matter density, the analysis will tend to converge towards the value of the free parameter that inherently satisfies this condition. Otherwise, the analysis risks yielding degenerate solutions.

\subsection{Analysis Methods and Strategies}

The objective of this analysis is to identify which features at the background level serve to alleviate the Hubble tension. Achieving this necessitates a multi-tiered approach, involving both various datasets and analysis techniques. Accordingly, we employ various combinations of data under stringent constraints applied to the models under consideration. Apart from $\Lambda$CDM, all other models in the comprehensive analysis include one or more additional free parameters.

There are two primary concerns regarding models that claim to resolve the Hubble tension. Regardless of the physical implications -- such as phantom crossing, negative dark energy densities, or other more exotic phenomena -- it is essential to develop a clear understanding of,
\begin{enumerate}
	\item How do the choice of data and priors influence the conclusions drawn?
	\item In what way do correlations between free parameters and data interact to shape these conclusions?
\end{enumerate}
Both of these questions are model-independent, provided that the models under consideration share similar or identical free parameters. In our case, once specific values are chosen for the additional free parameters, all models effectively possess the same free parameters as $\Lambda$CDM. Consequently, our focus will be on presenting results that emphasise the impact of data selection and prior assumptions.

Moreover, any estimation of free parameters performed using conventional MCMC chains yields the posterior distribution of the parameters alongside the model likelihood. Based on these outcomes, we will report the standard statistical measures, including the likelihood and various information criteria. Most importantly, we will examine the degree of tension between the estimated parameters and the globally accepted reference values.

\subsubsection{\textbf{Tension}} 

Conventionally, the tension between an estimated value, denoted as $\mathcal{A}_e \pm \sigma_{e}$, and a reference value, $\mathcal{A}_r \pm \sigma_{r}$,
\begin{equation}
\text{Tension}=\frac{|\mathcal{A}_e-\mathcal{A}_r|}{\sqrt{\sigma_{e}^2+\sigma_{r}^2}}
\end{equation}
When the estimates exhibit slight asymmetry in their upper and lower bounds, expressed as $\mathcal{A}_e + \sigma_{u} - \sigma_{l}$, we avoid using the average of these deviations. Instead, we use $\sigma_{u}$ if $\mathcal{A}_e$ is less than the reference value and $\sigma_{l}$ otherwise.

In this analysis, the tensions in the Hubble parameter are calculated with respect to $H_0^{\rm SH0ES} = 73.3 \pm 1.04$ for the SH0ES estimate and $H_0^{\rm CMB} = 67.4 \pm 0.5$ for the CMB-derived value. Additionally, we estimate the tension in the matter density parameter, using SH0ES-based estimates $\Omega_m^{\rm SH0ES} = 0.334 \pm 0.018$ and Planck 18-based estimates $\Omega_m^{\rm CMB} = 0.315 \pm 0.007$. The calculated tensions are presented in tables under the labels $\sigma_{H_0}^{\rm SH0ES}$, $\sigma_{H_0}^{\rm CMB}$, $\sigma_{\Omega_m}^{\rm SH0ES}$, and $\sigma_{\Omega_m}^{\rm CMB}$. This approach provides insights into how the correlation between $\Omega_m$ and $H_0$ influences the resolution of the Hubble tension and the role of data in shaping these parameter values.

\subsubsection{\textbf{Model Comparison}}
Given the variety of models available, it becomes necessary to select one over another, even though all may successfully account for the same observations. While a model that resolves the tension holds greater significance in many contexts, it is equally important to evaluate the cost of resolving the tension. Therefore, we employ several well-established methods to compare these models. It is crucial to note that the purpose of this manuscript is not to assert the superiority of one model over another; such an approach would fail to serve the intended purpose. In this context, we utilise,\\

\paragraph{\textbf{Log-Likelihood}}: Conventionally, authors report the minimum value of $\chi^2$, which is equivalent to providing the Log-Likelihood ($\ln {\cal L}(d|\theta)$) value. In this analysis, we report $-2\ln {\cal L}(d|\theta)$ instead of $\chi^2$ itself. Mathematically, we have,
\begin{align}
\label{eq:gaussianLikelihood}
\ln {\cal L}(d|\theta) = -\frac{1}{2}\sum_k^N 
\left\{\frac{\left[d_k-\mu_k(\theta)\right]^2}{\sigma_k^2}
+ \ln\left(2\pi\sigma^2_k\right)\right\}.
\end{align}
Here, $\ln {\cal L}(d|\theta)$ represents the log-likelihood of the parameter, $N$ is the number of data points, $d_k$ is the observed data point, $\mu_k(\theta)$ is the model prediction for the parameter $\theta$, and $\sigma_{k}$ is the standard deviation. Except for an additional constant term that depends entirely on the data, $-2\ln {\cal L}(d|\theta)$ is equivalent to $\chi^2$. This measure reflects the inherent errors within the data being used, rather than merely the number of data points. Moreover, for data sets with a covariance matrix, the expression becomes,
\begin{align}
\label{eq:gaussianLikelihoodCovariance}
\ln {\cal L}(d|\theta) = -\frac{1}{2} \left[ 
\Delta^T \Sigma^{-1} \Delta + \ln\left((2\pi)^N \det \Sigma\right) 
\right].
\end{align}
Here, $\Sigma$ represents the covariance matrix, $\Delta = d - \mu(\theta)$ is the residual vector, and $\Delta^T$ denotes its transpose. In such cases, we calculate $\chi^2$ using the formula $\chi^2 = \Delta^T \Sigma^{-1} \Delta$ or equivalently $\chi^2 = \frac{1}{2} \sum_k^N \frac{\left[d_k - \mu_k(\theta)\right]^2}{\sigma_k^2}$. Among various models applied to a fixed data set (or a specific combination of data sets), the model yielding the lowest $-2\ln {\cal L}(d|\theta)$ value is considered the best.\\

\paragraph{\textbf{Information Criterion}}: 
To address the challenges of model selection, H. Akaike introduced the Akaike Information Criterion (AIC) in 1974 \cite{1100705}, while G. Schwarz proposed the Bayesian Information Criterion (BIC) in 1978 \cite{10.1214/aos/1176344136}. These criteria have since become valuable tools for selecting appropriate models. The AIC focuses on the number of free parameters, whereas the BIC is grounded in Bayesian principles. Although similar, the BIC tends to favour models with fewer parameters, making it more suitable for model selection in many cases.
For situations where the parameter space is not well-defined, Spiegelhalter and colleagues proposed an alternative in 2002 \cite{10.1111/1467-9868.00353}. They suggested using the posterior mean deviance as a measure of fit and the individual contributions as a measure of complexity. This approach culminated in the development of the Deviance Information Criterion (DIC), which combines these elements for more generalised model comparison.\\

\noindent\textit{\textbf{Akaike Information Criterion}} (\textbf{AIC}):
\begin{align}
\mathrm{AIC} = 2N_p - 2 \ln \mathcal{L}_{\mathrm{max}},
\end{align}
where $ N_p $ is the number of model parameters and $ \ln \mathcal{L}_{\mathrm{max}} $ is the maximum log-likelihood.

\noindent\textit{\textbf{Bayesian Information Criterion}} (\textbf{BIC}):
\begin{align}
\mathrm{BIC} = N_p \ln N - 2 \ln \mathcal{L}_{\mathrm{max}},
\end{align}
where $ N $ is the number of data points.

\noindent\textit{\textbf{Deviance Information Criterion}} (\textbf{DIC}):
\begin{align}
\mathrm{DIC} &= \hat{D} + 2p_D.
\end{align}
Here, $\hat{D}$ is the deviance evaluated at the maximum likelihood estimate, $\bar{D}$ denotes the mean deviance, and $p_D = \bar{D}-\hat{D}$ indicates the effective number of parameters, also referred to as the model's complexity.

All these information criteria (ICs) share qualitative similarities, with BIC uniquely incorporating a multiplicative factor of $\log(N)/2$ applied to the number of free parameters, giving it an edge over AIC \cite{10.1214/aos/1176344136}. In many contexts, these ICs may appear nearly identical, especially when the prior is consistent. For instance, AIC and DIC often show striking numerical similarity, if not equivalence, under such conditions \cite{10.1111/1467-9868.00353}. The model with the lowest IC value is generally considered optimal.
However, our focus is not on using these criteria to elevate one model over another. Instead, we observe that the preference between models is usually marginal, with one model only slightly outperforming the other. Significant differences in preference are noted only when extensive datasets, such as Pantheon$^+$, are employed.

\section{What Solves the Hubble Tension?}

The discussion is twofold. First, we examine the impact of various datasets on each model's ability to alleviate the Hubble tension. Second, we analyse the intrinsic features of the models themselves that contribute to resolving the Hubble tension.

\subsection{Data Choices and Consequences}
In this section, we compare and contrast the PEDE and GOHDE models using various datasets. The primary objective is to analyse how the best-fit estimates are influenced by different combinations of data. A natural question arises: why revisit these models, which are already well-studied in the literature? The answer lies in the release of the DESI dataset, necessitating an update on the BAO $H(z)$ measurements. The new results, compared to their earlier counterparts, exhibit slight variations, prompting an investigation into their individual impacts on the analysis. Notably, a straightforward MCMC fitting -- performed with and without the data points at $z \sim 2.3$ -- revealed significant deviations in the best-fit values. Beyond the confidence intervals, the best-fit values appeared to exhibit tension within the datasets themselves.
\begin{figure*}
	\centering
	\includegraphics[width=0.32\textwidth]{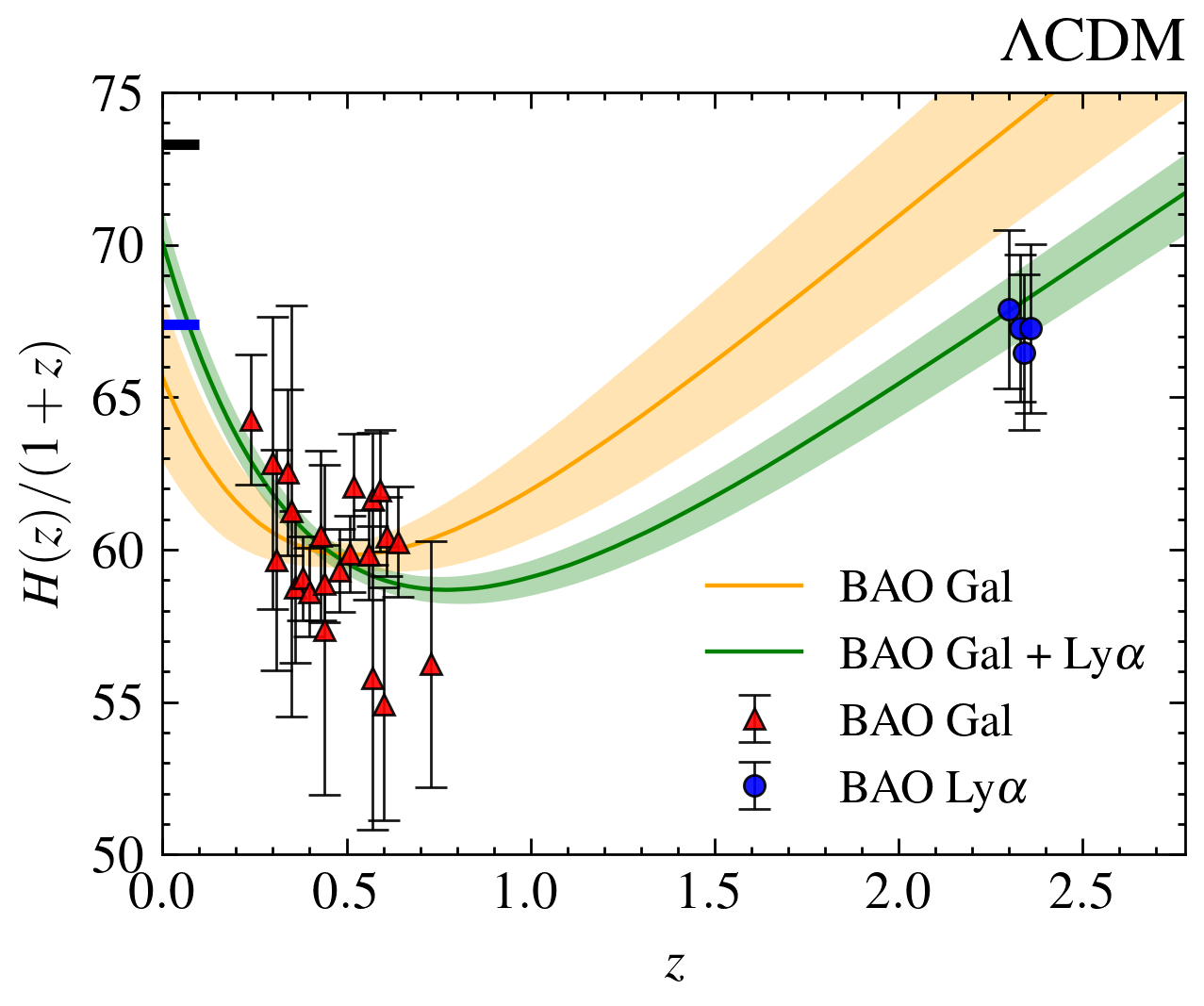}
	\includegraphics[width=0.32\textwidth]{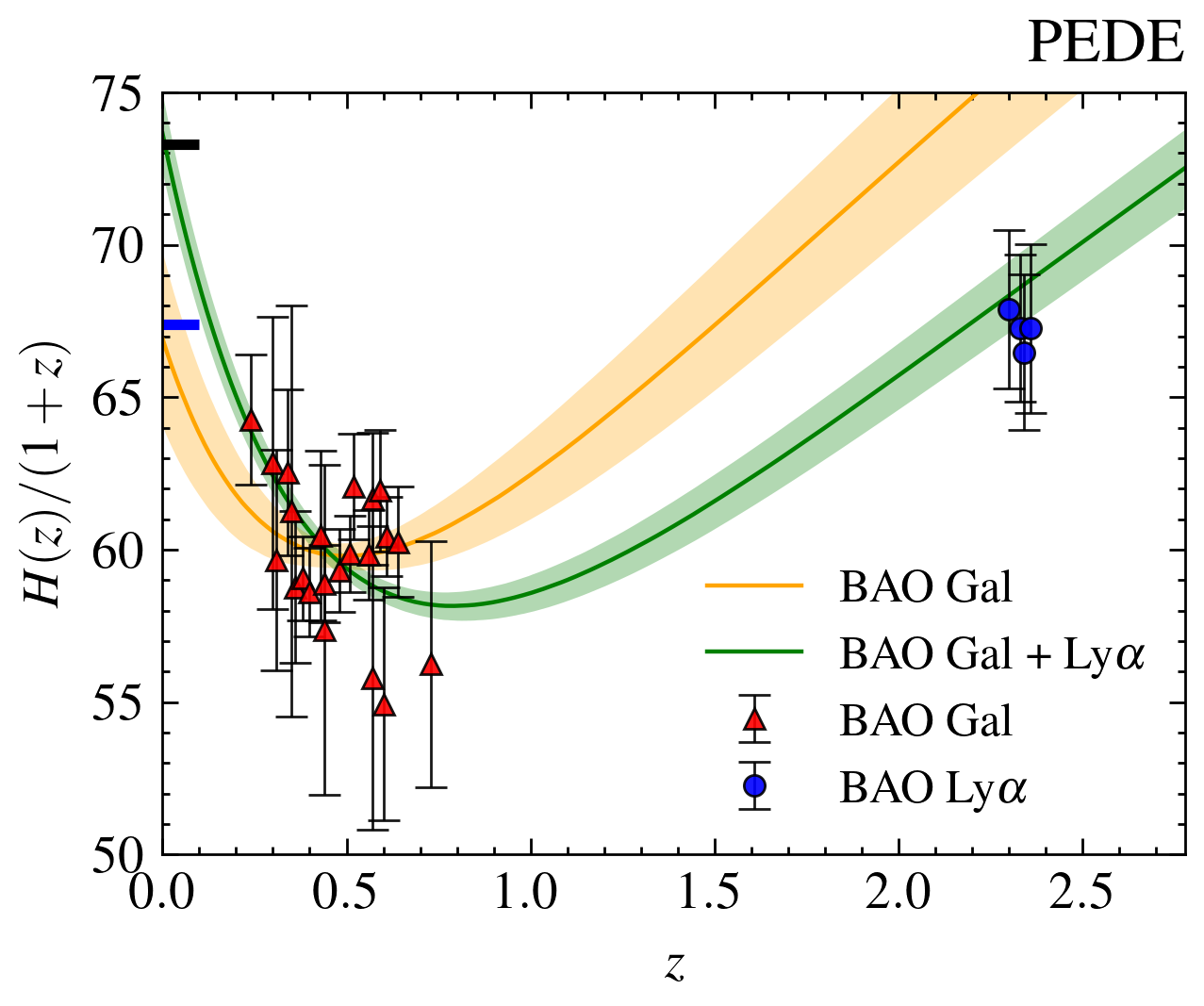}
	\includegraphics[width=0.32\textwidth]{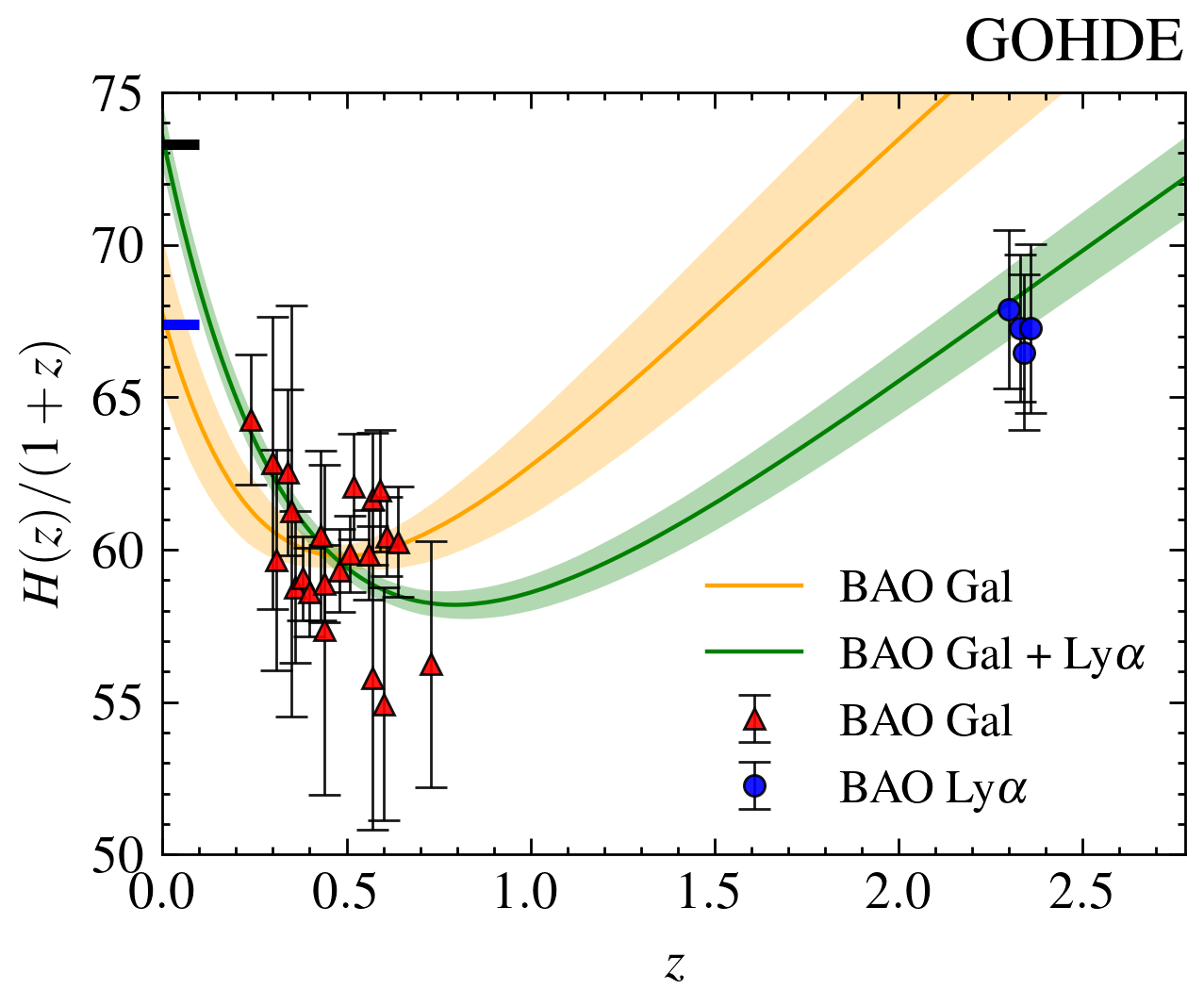}
	\caption{Comoving Hubble parameter as a function of redshift for the $\Lambda$CDM, PEDE ($\nu=1/\ln(10)$), and GOHDE ($\beta=1/3$) models, analysed using the BAO Gal data with and without the BAO Ly$\alpha$ contribution. The black and blue markers on the y-axis represent the SH0ES and Planck 18 estimates, respectively. The solid lines depict the median values from the MCMC-generated samples, while the shaded regions correspond to the 16th and 84th percentiles, representing the confidence intervals.\label{fig:HzBAOGalPlusMinusLya}}
\end{figure*}

In FIG. (\ref{fig:HzBAOGalPlusMinusLya}), we present the comoving Hubble parameter as a function of redshift ($z$). Each figure features two distinct curves, one derived using the BAO Gal data (orange) and the other incorporating the BAO Ly$\alpha$ data alongside the BAO Gal data (green). Across all models, a consistent pattern emerges when BAO Ly$\alpha$ data is added to the analysis -- specifically, a natural accommodation of the additional data points and an overall increase in the Hubble parameter.

\begin{figure}
	\centering
	\includegraphics[width=0.47\textwidth]{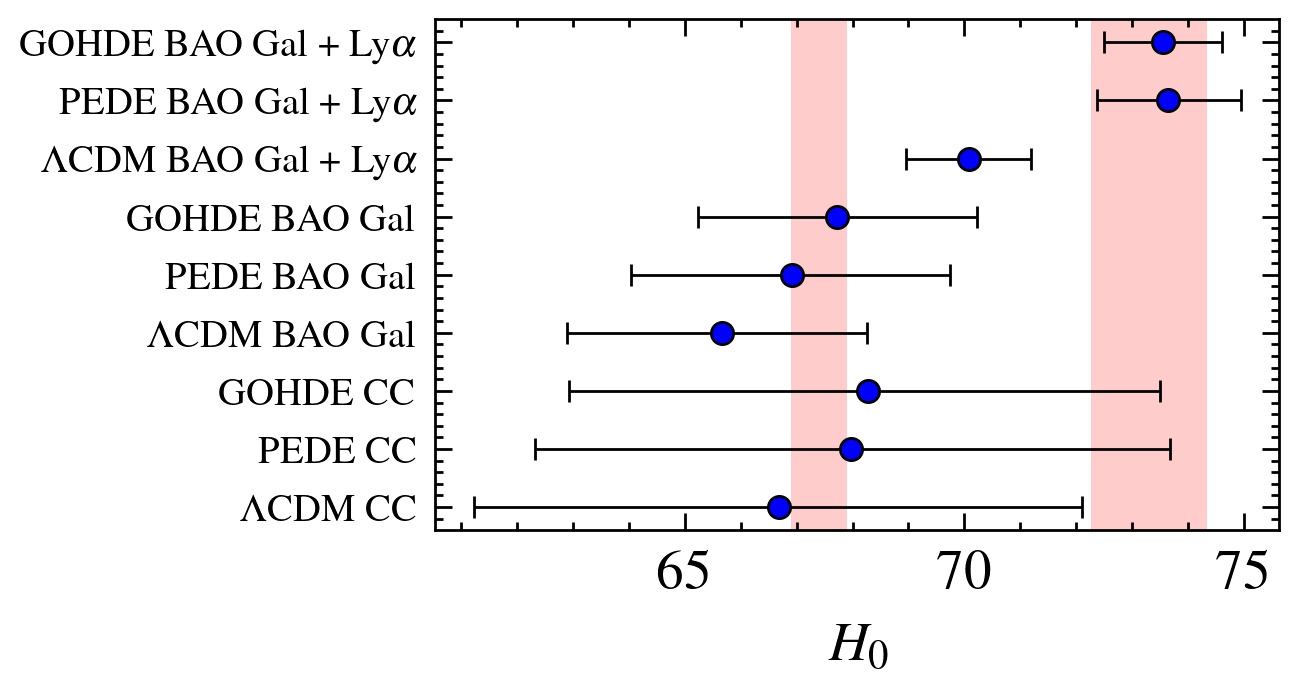}
	\caption{Comparison of Hubble parameter estimates for $\Lambda$CDM, PEDE, and GOHDE models using CC, BAO Gal, and BAO Gal+Ly$\alpha$ data sets. The BAO Ly$\alpha$ data set includes all points listed in TABLE (\ref{tab:BAOLyaAllOld}). The rose-shaded bands represent the Planck 18 estimate ($H_0^{\rm CMB} = 67.4 \pm 0.5$) and the SH0ES estimate ($H_0^{\rm SH0ES} = 73.3 \pm 1.04$), both in km$/$s$/$Mpc.\label{fig:CCBAOGALLYACom}}
\end{figure}

The BAO Gal data alone estimates the Hubble parameter close to the CMB predictions for all models: $\Lambda$CDM ($\sim 65.6\pm2.6$ km/s/Mpc), PEDE ($\sim 66.9\pm2.8$ km/s/Mpc), and GOHDE ($\sim 67.7\pm2.5$ km/s/Mpc). Similar estimates are obtained when using CC data, albeit with larger error bars. Due to these significant error margins, the CC data alone cannot indicate a tension with the SH0ES value. However, the BAO Gal data displays a pronounced tension with the SH0ES value. Interestingly, both the PEDE and GOHDE models outperform $\Lambda$CDM in alleviating the $H_0$ tension. While this improvement is promising, it is not substantial enough to be considered a resolution. Additionally, though the differences are minor, all information criteria show a preference for the PEDE and GOHDE models over $\Lambda$CDM. Between PEDE and GOHDE, no significant differences are observed when using the CC and BAO Gal data. Detailed results are provided in TABLE (\ref{tab:CCBAOGAL}).
\begin{table*}
	\centering
	\renewcommand{\arraystretch}{1.3}
	\begin{ruledtabular}
		\begin{tabular}{lcccccccccc}
			Model & $\Omega_m$ or $\tilde{\Omega}_m$ & $H_0$ & $\sigma_{H_0}^{\rm SH0ES}$ & $\sigma_{H_0}^{\rm CMB}$ & $\sigma_{\Omega_m}^{\rm SH0ES}$ & $\sigma_{\Omega_m}^{\rm CMB}$ & $-2\ln {\cal L}$ & AIC & BIC & DIC \\
			\hline
			\multicolumn{11}{c}{Data: CC} \\
			\hline
			$\Lambda$CDM & $0.334^{+0.081}_{-0.063}$ & $66.679^{+5.418}_{-5.456}$ & $1.2\sigma$ & $0.1\sigma$ & $0.0\sigma$ & $0.3\sigma$ & $248.7$ & $252.7$ & $265.9$ & $252.7$ \\
			PEDE& $0.340^{+0.078}_{-0.060}$ & $67.973^{+5.709}_{-5.643}$ & $0.9\sigma$ & $0.1\sigma$ & $0.1\sigma$ & $0.4\sigma$ & $248.5$ & $252.5$ & $265.8$ & $252.7$ \\
			GOHDE & $0.342^{+0.080}_{-0.056}$ & $68.276^{+5.216}_{-5.341}$ & $0.9\sigma$ & $0.2\sigma$ & $0.09\sigma$ & $0.33\sigma$ & $248.5$ & $252.5$ & $265.7$ & $252.6$ \\
			\hline
			\multicolumn{11}{c}{Data: BAO Gal} \\
			\hline
			$\Lambda$CDM & $0.366^{+0.072}_{-0.060}$ & $65.657^{+2.608}_{-2.756}$ & $2.7\sigma$ & $0.7\sigma$ & $0.5\sigma$ & $0.8\sigma$ & $108.7$ & $112.7$ & $126.0$ & $112.9$ \\
			PEDE & $0.381^{+0.066}_{-0.057}$ & $66.916^{+2.823}_{-2.883}$ & $2.1\sigma$ & $0.2\sigma$ & $0.8\sigma$ & $1.1\sigma$ & $108.4$ & $112.4$ & $125.6$ & $112.5$ \\
			GOHDE & $0.382^{+0.067}_{-0.055}$ & $67.728^{+2.496}_{-2.497}$ & $2.1\sigma$ & $0.1\sigma$ & $0.69\sigma$ & $1.00\sigma$ & $108.3$ & $112.3$ & $125.5$ & $112.3$ \\
		\end{tabular}
		\label{tab:parameter_constraints}
	\end{ruledtabular}
\caption{Best-fit values of $H_0$ (in km$/$s$/$Mpc) and $\Omega_m$ ($\tilde{\Omega}_m$) for the PEDE and GOHDE models using CC and BAO Gal data sets, along with the corresponding $-2\ln {\cal L}$, AIC, BIC, and DIC values. The table also includes the tension between the best-fit estimates and the SH0ES and CMB values for these parameters.\label{tab:CCBAOGAL}}
\end{table*}
Rather than modifying the model, let us examine the impact of incorporating the BAO Ly$\alpha$ data points. Here, we include all the data listed in TABLE (\ref{tab:BAOLyaAllOld}). The inclusion of these points resulted in a notable shift in the $H_0$ estimates for all models, with PEDE and GOHDE demonstrating a significant reduction in the Hubble tension. This effect is clearly illustrated in FIG. (\ref{fig:CCBAOGALLYACom}).
The best fit estimates are,
\begin{align*}
\text{PEDE:} ~& H_0 = 73.65 \pm 1.29 \, \text{km$/$s$/$Mpc}, \\&\Omega_m = 0.25 \pm 0.02, \\
\text{GOHDE:} ~& H_0 = 73.55 \pm 1.06 \, \text{km$/$s$/$Mpc},  \\&\Omega_m = 0.25 \pm 0.015, \\
\Lambda\text{CDM:} ~& H_0 = 70.09 \pm 1.1 \, \text{km$/$s$/$Mpc}, \\&\Omega_m = 0.26 \pm 0.02.
\end{align*}
Although $\Lambda$CDM improved its estimates, both PEDE and GOHDE successfully resolved the tension with the SH0ES $H_0$ estimate. However, this resolution introduced a significant $3-4\sigma$ tension in $\Omega_{m}$. While the BAO Gal data initially suggested a lower $H_0$ value and a higher $\Omega_{m}$ value, it maintained consistency with the SH0ES and CMB estimates for $\Omega_{m}$. This dynamic shifted upon the inclusion of just four additional data points from BAO Ly$\alpha$, reversing the trends. The resolution of the Hubble tension at the background level is closely tied to the correlation between $H_0$ and $\Omega_{m}$. Despite these complexities, both PEDE and GOHDE have demonstrated their ability to alleviate the tension at the background level, as shown in FIG. (\ref{fig:HzBAOGalPlusMinusLya}). Furthermore, the turning point in the comoving Hubble flow has shifted toward earlier times, yet remains within the predicted acceptable range. 

The key question is, why does this happen? These data points are tightly constrained, with very small error bars, so even just four of them can significantly influence the likelihood. But why does $H_0$ increase? This is related to the fiducial cosmology assumed when estimating these points. In \cite{refId0Busca, Font-Ribera_2014, refId0Delubac, refId0Bautista}, a cosmology with $\Omega_{m} = 0.27$ was assumed—about $1\sigma$ away from the current estimates. Therefore, the matter density values obtained remain consistent with this fiducial cosmology. Notably, with $\Lambda$CDM, the estimates align perfectly with the analysis protocols of these references, while PEDE and GOHDE emerge as two models capable of resolving the Hubble tension.

\begin{figure*}[t]
	\centering
	\includegraphics[width=0.32\textwidth]{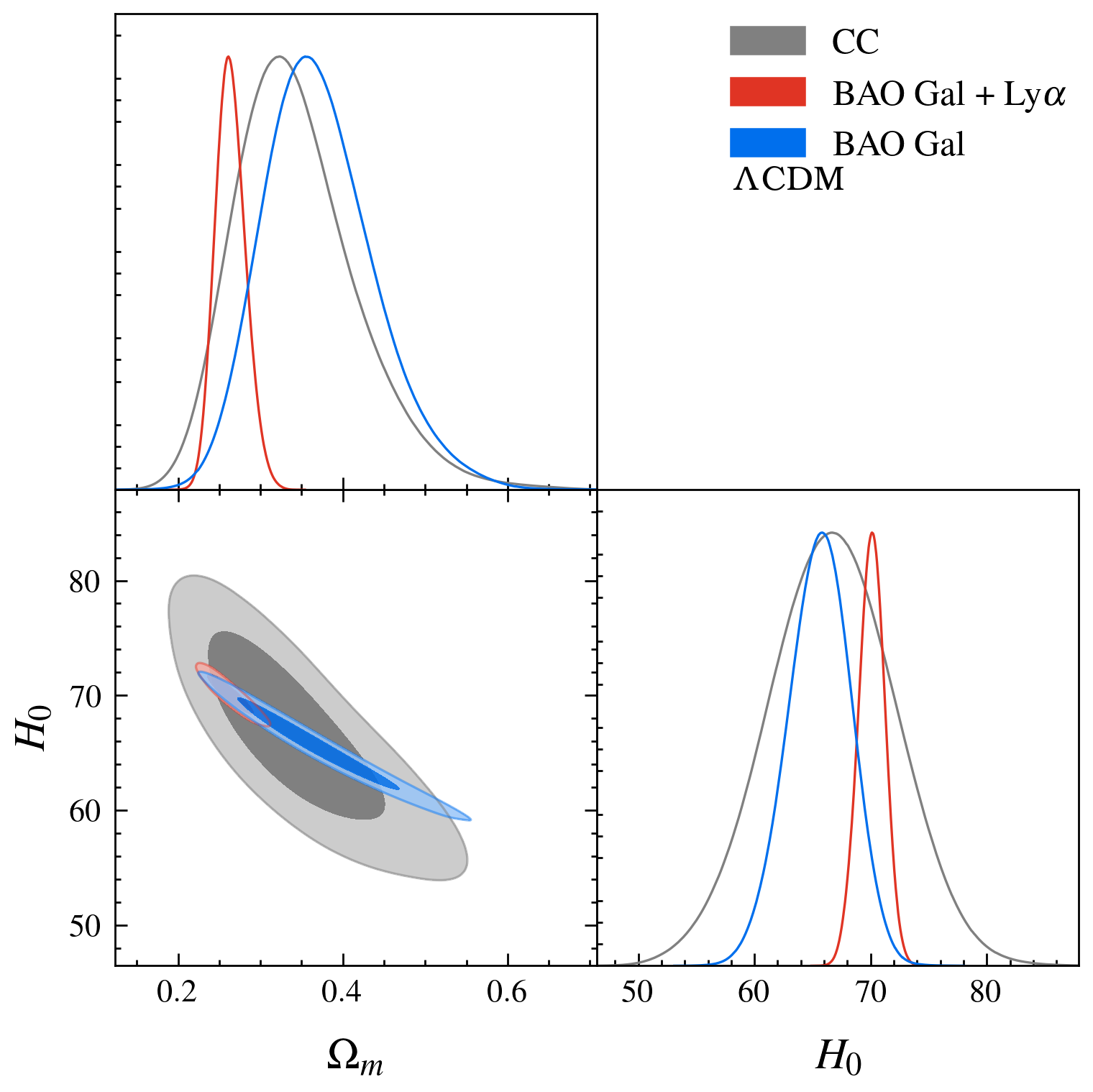}
	\includegraphics[width=0.32\textwidth]{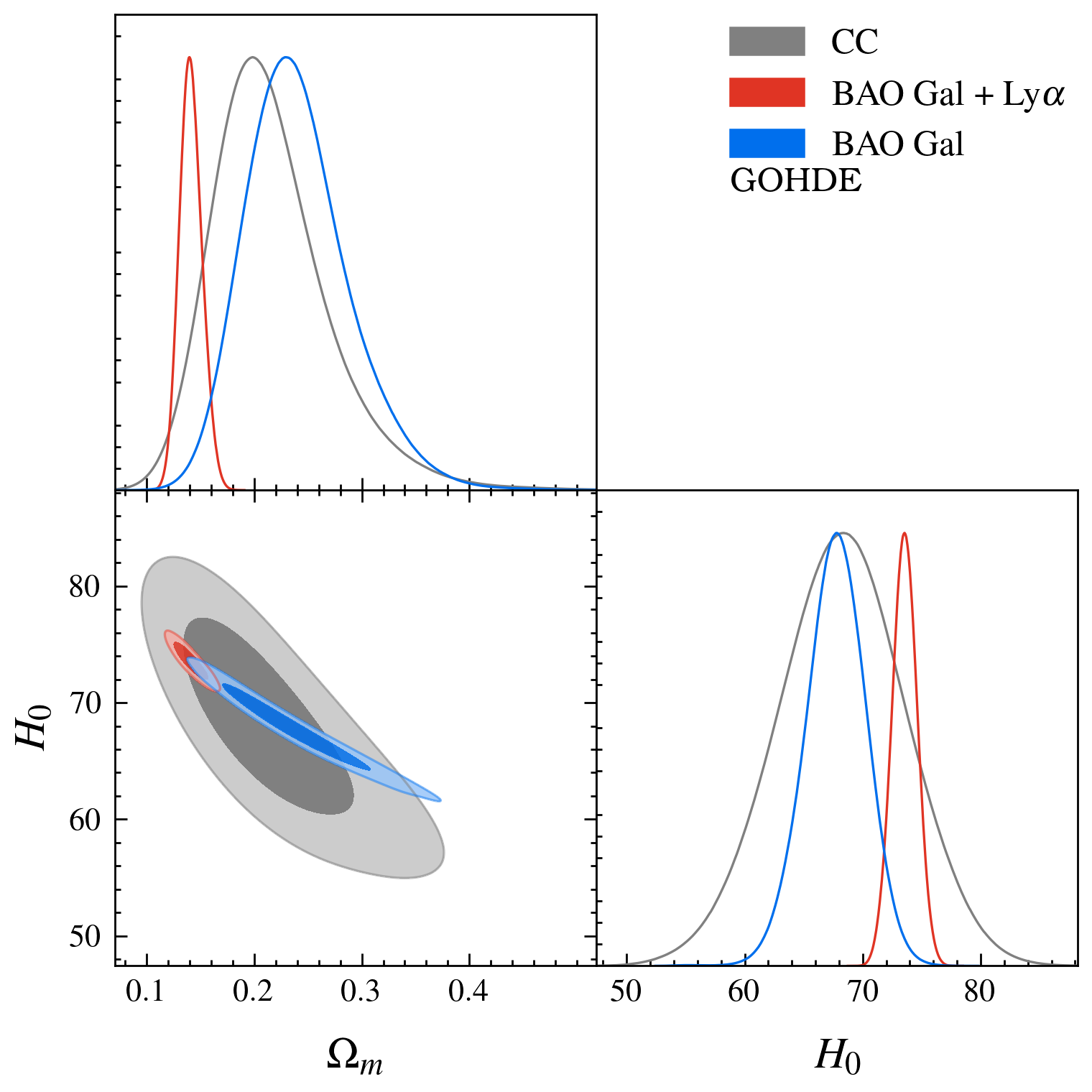}
	\includegraphics[width=0.32\textwidth]{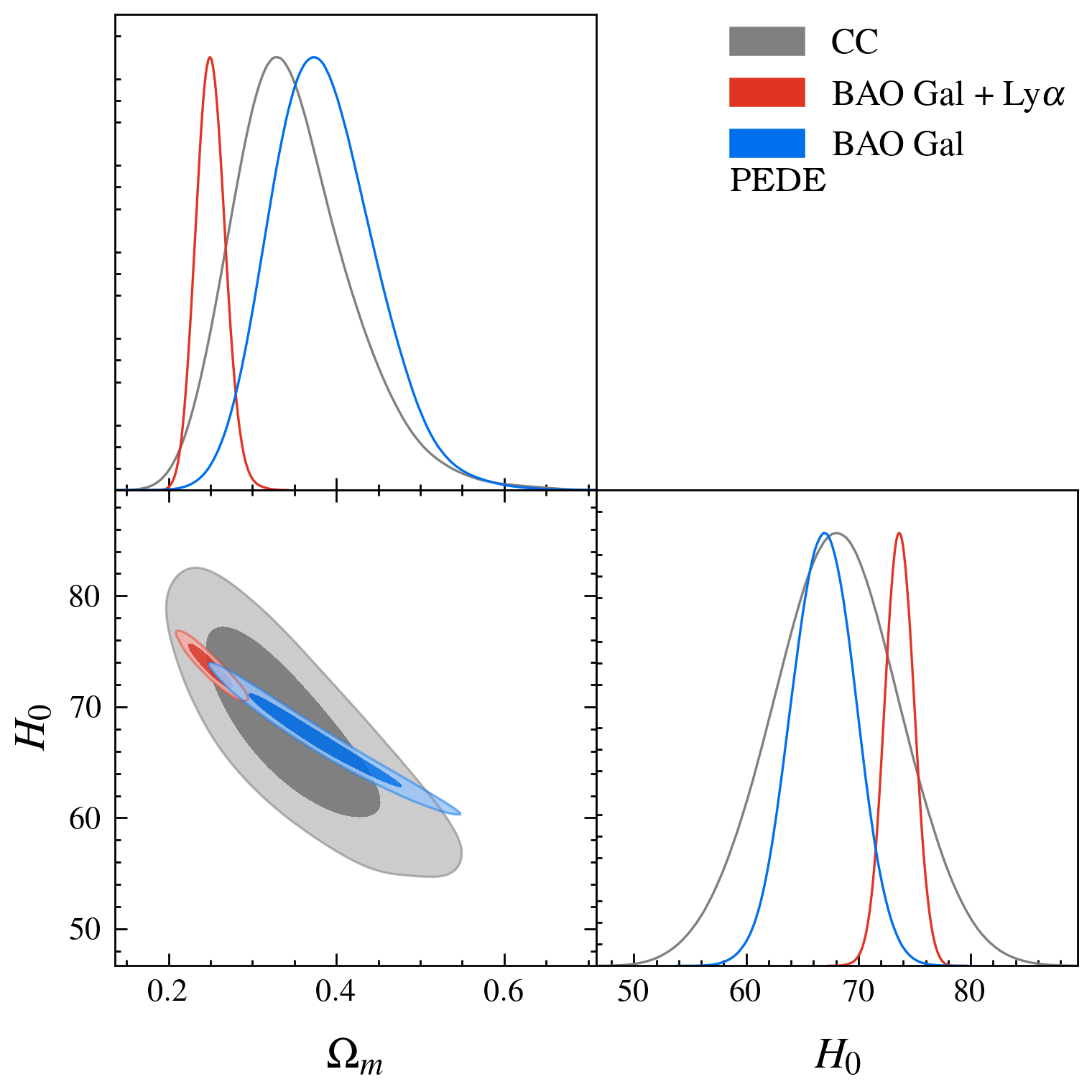}
	\caption{Corner plots for $\Omega_{m}$ and $H_0$ for $\Lambda$CDM, PEDE and GOHDE  models for various datasets indicating the  variation in the confidence interval.  Here, for GOHDE model we are using the $\Omega_{m}$ instead of the effective matter density $\tilde{\Omega}_m$.  \label{fig:baselcdmpedegohde}}
\end{figure*}

Therefore, we conclude that models like PEDE and GOHDE inherently possess the capability to resolve the Hubble tension. We will examine the specific features of these models in more detail later. The primary reason these models alleviate the Hubble tension when including the BAO Ly$\alpha$ data is the constraints on $\Omega_{m}$ imposed by those data points within the assumed fiducial cosmology. This explains why, despite these data points adopting a drag radius consistent with the CMB, they yield a higher $H_0$ value. The confidence contours for this analysis are presented in FIG. (\ref{fig:baselcdmpedegohde}), where the improvement and shift in parameter estimates are observable. In FIG. (\ref{fig:baselcdmpedegohde}), note that $\Omega_{m}$ is plotted instead of $\tilde{\Omega}_m$ for the GOHDE model; the nonlinear relationship between these parameters should be kept in mind when interpreting the figures.

Similarly, one might ask what happens when using DESI DR1 and DR2 BAO Ly$\alpha$ data points. In this case, we also have four points, but instead of using them all simultaneously, we analyse them individually. Specifically, there are two points from each survey: one from the Ly$\alpha$–Ly$\alpha$ forest auto-correlation study, and another including additional quasar (QSO) cross-correlation data. These points exhibit mild tension due to a lower high-redshift expansion rate \cite{akarsu2025dynamical}. The QSO data itself contributes to these tensions, which have been observed across various cosmological models, including GOHDE \cite{Manoharan2024}. We label these points as Ly$\alpha$–QSO for the QSO cross-correlation data, and Ly$\alpha$–QSO$+$ when combining Ly$\alpha$–Ly$\alpha$ auto-correlations with QSO data. We then consider estimates from SDSS and DESI separately, both with and without CC data. The resulting best-fit values and associated metrics are presented in TABLES (\ref{tab:SDSS}) and (\ref{tab:DESI}).
\begin{table*}
	\renewcommand{\arraystretch}{1.3}
	\centering
	\begin{ruledtabular}
		\begin{tabular}{lcccccccccc}
			Model&$\Omega_m$ or $\tilde{\Omega}_m$ & $H_0$ & $\sigma_{H_0}^{\rm SH0ES}$ & $\sigma_{H_0}^{\rm CMB}$ & $\sigma_{\Omega_m}^{\rm SH0ES}$ & $\sigma_{\Omega_m}^{\rm CMB}$ & $-2\ln {\cal L}$ & AIC & BIC & DIC \\
			\hline
			\multicolumn{11}{c}{BAO-SDSS-Ly$\alpha$-QSO} \\
			\hline
			$\Lambda$CDM & $0.289^{+0.034}_{-0.031}$ & $68.866^{+1.591}_{-1.662}$ & $2.3\sigma$ & $0.8\sigma$ & $1.2\sigma$ & $0.7\sigma$ & $117.1$ & $121.1$ & $134.2$ & $121.0$ \\
			PEDE & $0.287^{+0.034}_{-0.029}$ & $71.565^{+1.817}_{-1.962}$ & $0.8\sigma$ & $2.1\sigma$ & $1.2\sigma$ & $0.8\sigma$ & $119.3$ & $123.3$ & $136.5$ & $123.3$ \\
			GOHDE & $0.280^{+0.033}_{-0.028}$ & $72.044^{+1.481}_{-1.597}$ & $0.7\sigma$ & $2.8\sigma$ & $1.6\sigma$ & $1.2\sigma$ & $119.3$ & $123.3$ & $136.5$ & $123.3$ \\
			\hline
			\multicolumn{11}{c}{BAO-SDSS-Ly$\alpha$-QSO + CC} \\
			\hline
			$\Lambda$CDM & $0.299^{+0.030}_{-0.028}$ & $68.405^{+1.471}_{-1.518}$ & $2.7\sigma$ & $0.6\sigma$ & $1.0\sigma$ & $0.5\sigma$ & $366.1$ & $370.1$ & $383.2$ & $370.1$ \\
			PEDE & $0.297^{+0.030}_{-0.026}$ & $70.966^{+1.588}_{-1.731}$ & $1.2\sigma$ & $2.0\sigma$ & $1.1\sigma$ & $0.6\sigma$ & $368.3$ & $372.3$ & $385.5$ & $372.2$ \\
			GOHDE & $0.291^{+0.028}_{-0.025}$ & $71.496^{+1.384}_{-1.360}$ & $1.0\sigma$ & $2.8\sigma$ & $1.4\sigma$ & $0.9\sigma$ & $368.3$ & $372.3$ & $385.5$ & $372.3$ \\
			\hline
			\multicolumn{11}{c}{BAO-SDSS-Ly$\alpha$-QSO$^+$} \\
			\hline
			$\Lambda$CDM & $0.273^{+0.019}_{-0.018}$ & $69.616^{+1.152}_{-1.116}$ & $2.4\sigma$ & $1.8\sigma$ & $2.3\sigma$ & $2.1\sigma$ & $122.5$ & $126.5$ & $139.7$ & $126.5$ \\
			PEDE & $0.258^{+0.019}_{-0.017}$ & $73.167^{+1.301}_{-1.370}$ & $0.1\sigma$ & $4.0\sigma$ & $2.9\sigma$ & $2.9\sigma$ & $125.7$ & $129.7$ & $142.9$ & $129.7$ \\
			GOHDE & $0.255^{+0.017}_{-0.015}$ & $73.201^{+1.019}_{-1.046}$ & $0.1\sigma$ & $5.0\sigma$ & $3.4\sigma$ & $3.6\sigma$ & $125.3$ & $129.3$ & $142.4$ & $129.2$ \\
			\hline
			\multicolumn{11}{c}{BAO-SDSS-Ly$\alpha$-QSO$^+$ + CC} \\
			\hline
			$\Lambda$CDM & $0.277^{+0.018}_{-0.017}$ & $69.390^{+1.068}_{-1.111}$ & $2.6\sigma$ & $1.6\sigma$ & $2.2\sigma$ & $1.9\sigma$ & $372.0$ & $376.0$ & $389.2$ & $375.9$ \\
			PEDE & $0.265^{+0.017}_{-0.017}$ & $72.696^{+1.264}_{-1.235}$ & $0.4\sigma$ & $4.0\sigma$ & $2.7\sigma$ & $2.6\sigma$ & $375.9$ & $379.9$ & $393.0$ & $379.9$ \\
			GOHDE & $0.261^{+0.017}_{-0.015}$ & $72.823^{+1.008}_{-1.046}$ & $0.3\sigma$ & $4.7\sigma$ & $3.1\sigma$ & $3.2\sigma$ & $375.4$ & $379.4$ & $392.6$ & $379.5$ \\
		\end{tabular}
	\end{ruledtabular}
\caption{Best-fit values of $H_0$ (in km$/$s$/$Mpc)  and $\Omega_m$ (or $\tilde{\Omega}_m$) derived from SDSS data for the $\Lambda$CDM, PEDE, and GOHDE models, along with their corresponding $-2\ln {\cal L}$, AIC, BIC, and DIC statistics. The table also shows the tension of these fits relative to SH0ES and CMB estimates for the parameters.\label{tab:SDSS}}
\end{table*}
\begin{table*}
	\renewcommand{\arraystretch}{1.3}
	\centering
	\begin{ruledtabular}
		\begin{tabular}{lcccccccccc}
			Model&$\Omega_m$ & $H_0$ & $\sigma_{H_0}^{\rm SH0ES}$ & $\sigma_{H_0}^{\rm CMB}$ & $\sigma_{\Omega_m}^{\rm SH0ES}$ & $\sigma_{\Omega_m}^{\rm CMB}$ & $-2\ln {\cal L}$ & AIC & BIC & DIC \\
			\hline
			\multicolumn{11}{c}{BAO-DESI-Ly$\alpha$-QSO} \\
			\hline
			$\Lambda$CDM& $0.329^{+0.026}_{-0.024}$ & $67.107^{+1.286}_{-1.329}$ & $3.7\sigma$ & $0.2\sigma$ & $0.2\sigma$ & $0.6\sigma$ & $114.0$ & $118.0$ & $131.2$ & $118.0$ \\
			PEDE& $0.313^{+0.026}_{-0.023}$ & $70.160^{+1.484}_{-1.499}$ & $1.7\sigma$ & $1.7\sigma$ & $0.7\sigma$ & $0.1\sigma$ & $114.9$ & $118.9$ & $132.1$ & $118.9$ \\
			GOHDE& $0.307^{+0.023}_{-0.020}$ & $70.842^{+1.215}_{-1.212}$ & $1.5\sigma$ & $2.6\sigma$ & $1.0\sigma$ & $0.4\sigma$ & $115.0$ & $119.0$ & $132.2$ & $119.1$ \\
			\hline
			\multicolumn{11}{c}{BAO-DESI-Ly$\alpha$-QSO + CC} \\
			\hline
			$\Lambda$CDM& $0.328^{+0.025}_{-0.022}$ & $67.167^{+1.166}_{-1.255}$ & $3.9\sigma$ & $0.2\sigma$ & $0.2\sigma$ & $0.6\sigma$ & $362.6$ & $366.6$ & $379.8$ & $366.7$ \\
			PEDE & $0.315^{+0.023}_{-0.021}$ & $70.008^{+1.318}_{-1.362}$ & $2.0\sigma$ & $1.8\sigma$ & $0.6\sigma$ & $0.0\sigma$ & $363.5$ & $367.5$ & $380.7$ & $367.5$ \\
			GOHDE & $0.308^{+0.021}_{-0.020}$ & $70.749^{+1.171}_{-1.143}$ & $1.6\sigma$ & $2.7\sigma$ & $1.0\sigma$ & $0.3\sigma$ & $363.6$ & $367.6$ & $380.9$ & $367.8$ \\
			\hline
			\multicolumn{11}{c}{BAO-DESI-Ly$\alpha$-QSO$^+$} \\
			\hline
			$\Lambda$CDM& $0.313^{+0.015}_{-0.014}$ & $67.796^{+0.994}_{-1.011}$ & $3.8\sigma$ & $0.4\sigma$ & $0.9\sigma$ & $0.1\sigma$ & $118.5$ & $122.5$ & $135.7$ & $122.5$ \\
			PEDE & $0.291^{+0.014}_{-0.013}$ & $71.316^{+1.053}_{-1.119}$ & $1.3\sigma$ & $3.2\sigma$ & $1.9\sigma$ & $1.5\sigma$ & $120.1$ & $124.1$ & $137.2$ & $124.0$ \\
			GOHDE & $0.288^{+0.013}_{-0.012}$ & $71.702^{+0.898}_{-0.955}$ & $1.2\sigma$ & $4.0\sigma$ & $2.2\sigma$ & $2.0\sigma$ & $120.1$ & $124.1$ & $137.2$ & $124.1$ \\
			\hline
			\multicolumn{11}{c}{BAO-DESI-Ly$\alpha$-QSO$^+$ + CC} \\
			\hline
			$\Lambda$CDM& $0.314^{+0.015}_{-0.014}$ & $67.763^{+0.987}_{-0.950}$ & $3.9\sigma$ & $0.3\sigma$ & $0.9\sigma$ & $0.1\sigma$ & $367.2$ & $371.2$ & $384.3$ & $371.2$ \\
			PEDE & $0.294^{+0.014}_{-0.013}$ & $71.126^{+1.061}_{-1.068}$ & $1.5\sigma$ & $3.2\sigma$ & $1.8\sigma$ & $1.4\sigma$ & $369.0$ & $373.0$ & $386.2$ & $373.0$ \\
			GOHDE & $0.289^{+0.013}_{-0.012}$ & $71.601^{+0.928}_{-0.898}$ & $1.2\sigma$ & $4.1\sigma$ & $2.1\sigma$ & $1.9\sigma$ & $369.0$ & $373.0$ & $386.3$ & $373.1$ \\
		\end{tabular}
	\end{ruledtabular}
\caption{Best-fit values of $H_0$ (in km$/$s$/$Mpc)  and $\Omega_m$ ($\tilde{\Omega}_m$) obtained from DESI DR1 and DR2 data for the $\Lambda$CDM, PEDE, and GOHDE models, along with the corresponding $-2\ln {\cal L}$, AIC, BIC, and DIC values. The table also includes the tension of these fits relative to SH0ES and CMB estimates for these parameters. \label{tab:DESI}}
\end{table*}

\begin{figure*}
	\centering
	\includegraphics[width=0.47\textwidth]{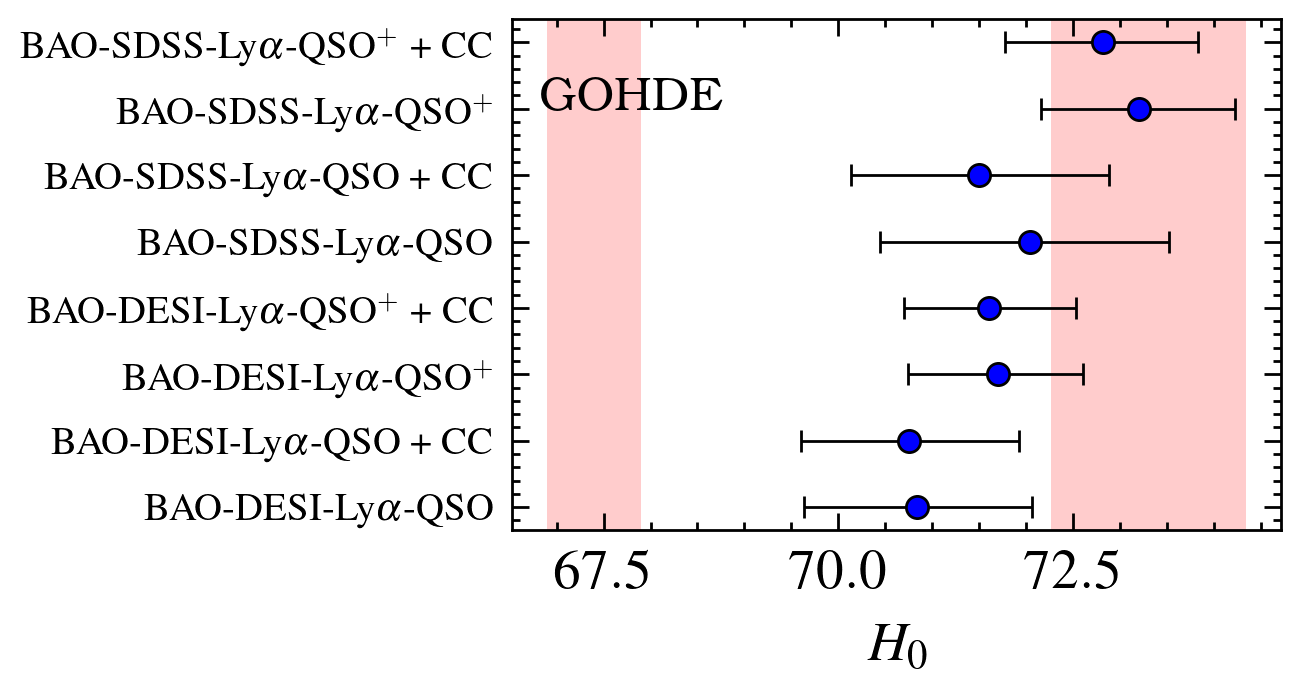}
	\includegraphics[width=0.47\textwidth]{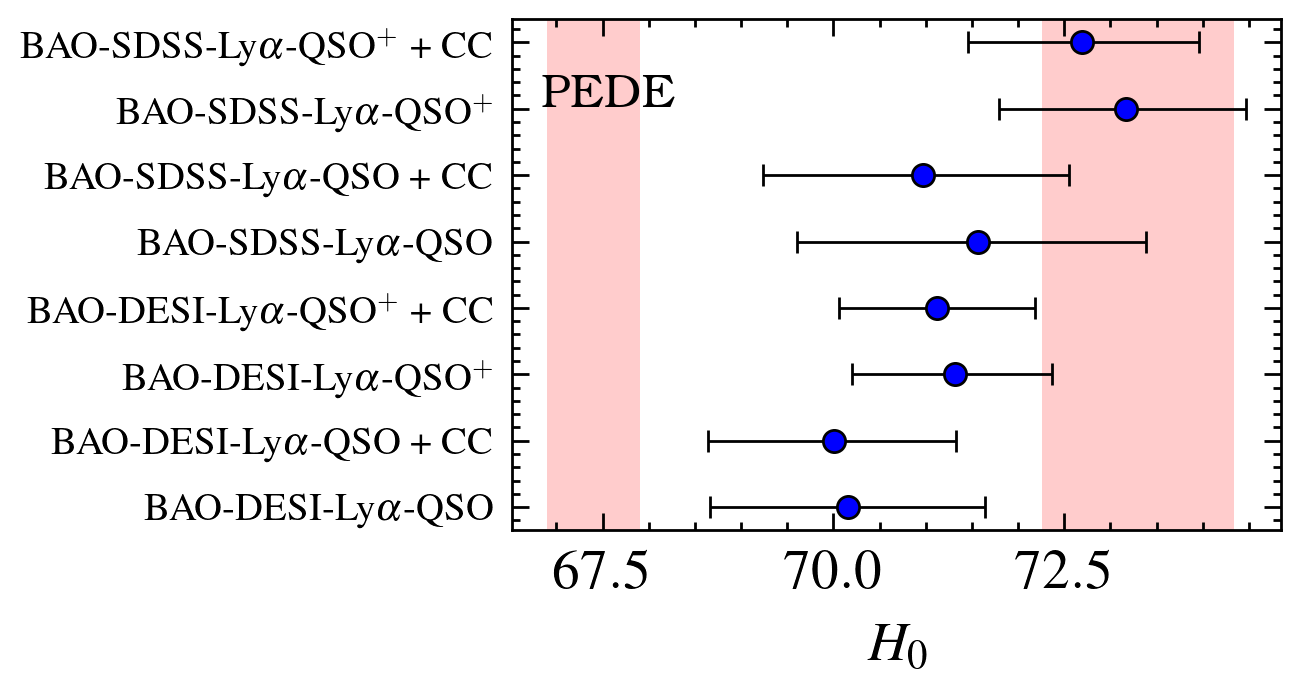}\\
	\includegraphics[width=0.47\textwidth]{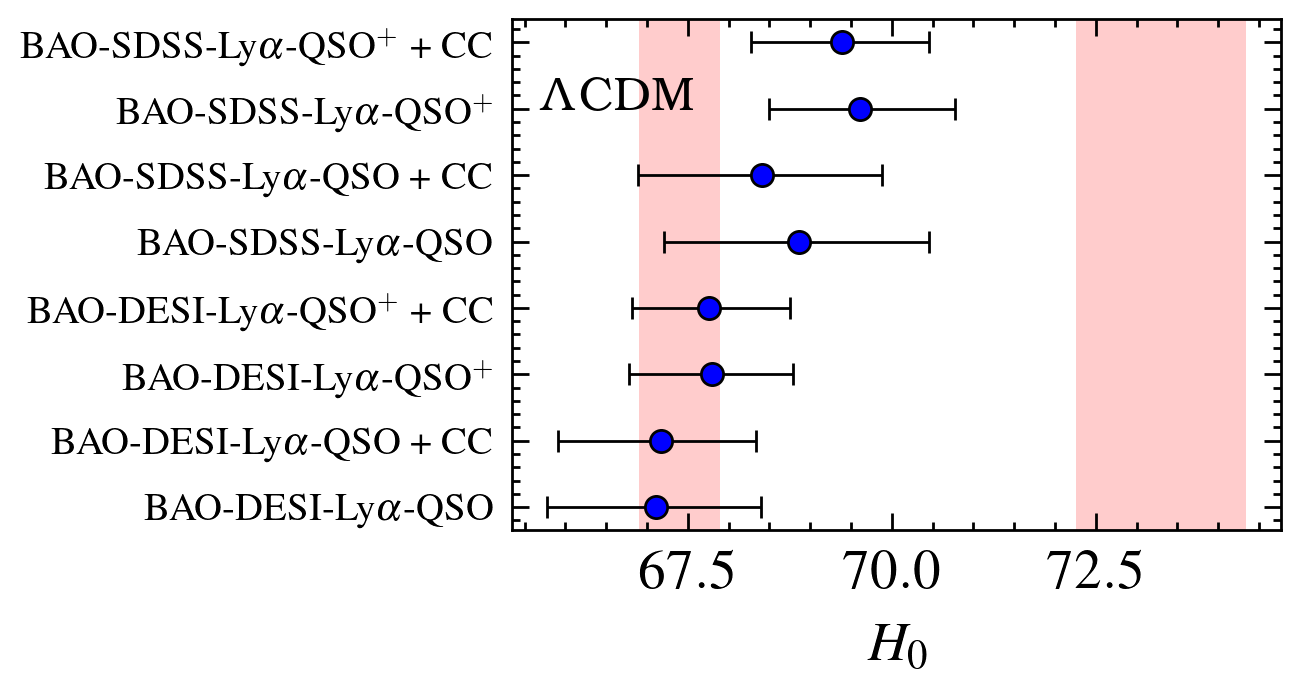}
	\caption{Comparison of Hubble parameter estimates for $\Lambda$CDM, PEDE and GOHDE models using various BAO data sets, with and without CC data. The rose bands represent the Planck 18 estimate ($H_0^{\rm CMB} = 67.4 \pm 0.5$) and the SH0ES estimate ($H_0^{\rm SH0ES} = 73.3 \pm 1.04$) in km$/$s$/$Mpc. \label{fig:DESISDSSlcdmPEDEGOHDE}}
\end{figure*}

For a clearer understanding of the table, one can refer to FIG. (\ref{fig:DESISDSSlcdmPEDEGOHDE}) for the $H_0$ estimates using the $\Lambda$CDM, PEDE and GOHDE models. 
Across all cases, the trends generated by the data remain consistent; however, PEDE and GOHDE demonstrate greater effectiveness in addressing the Hubble tension. Notably, the SDSS-derived estimates are more effective in resolving the tension, while the inclusion of Ly$\alpha$ auto-correlation data shows a similar and enhanced capacity to mitigate the Hubble tension. Adding CC data slightly worsens the estimates, though the error margins remain largely unaffected due to the collective BAO Gal data. For the $\Lambda$CDM model, the $H_0$ values continue to exhibit tension with SH0ES.

\begin{figure*}
	\centering
	\includegraphics[width=0.47\textwidth]{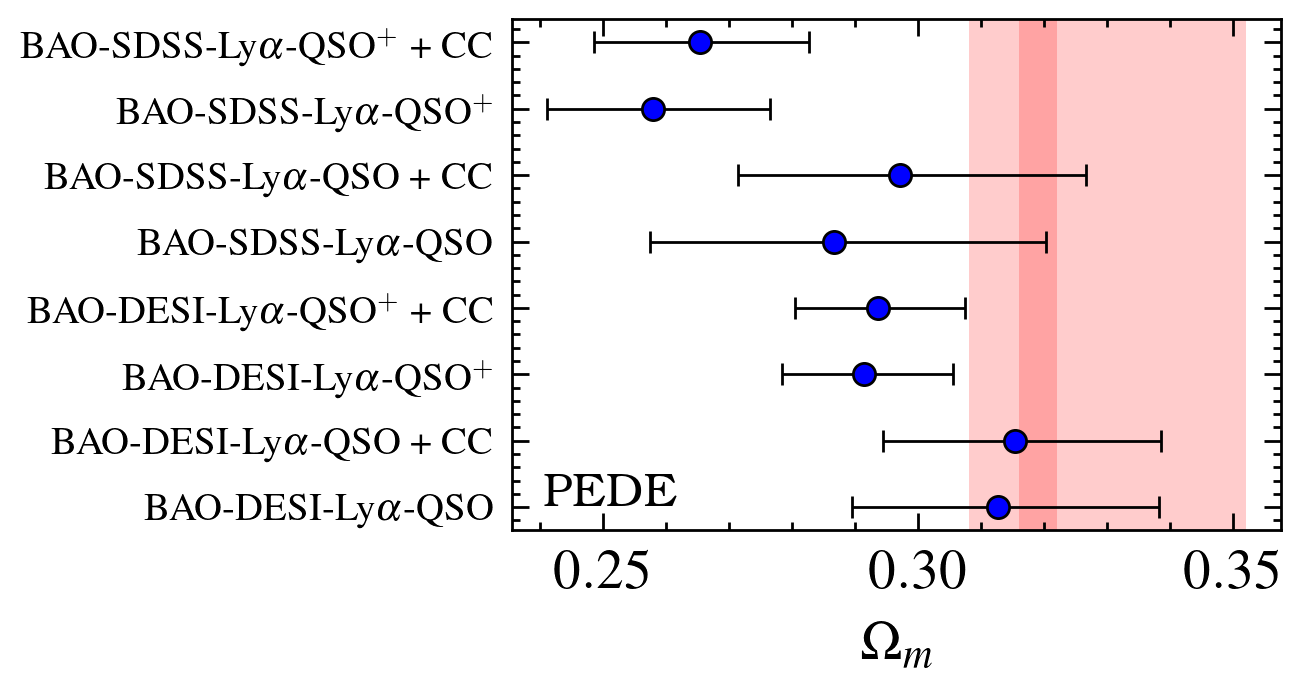}
	\includegraphics[width=0.47\textwidth]{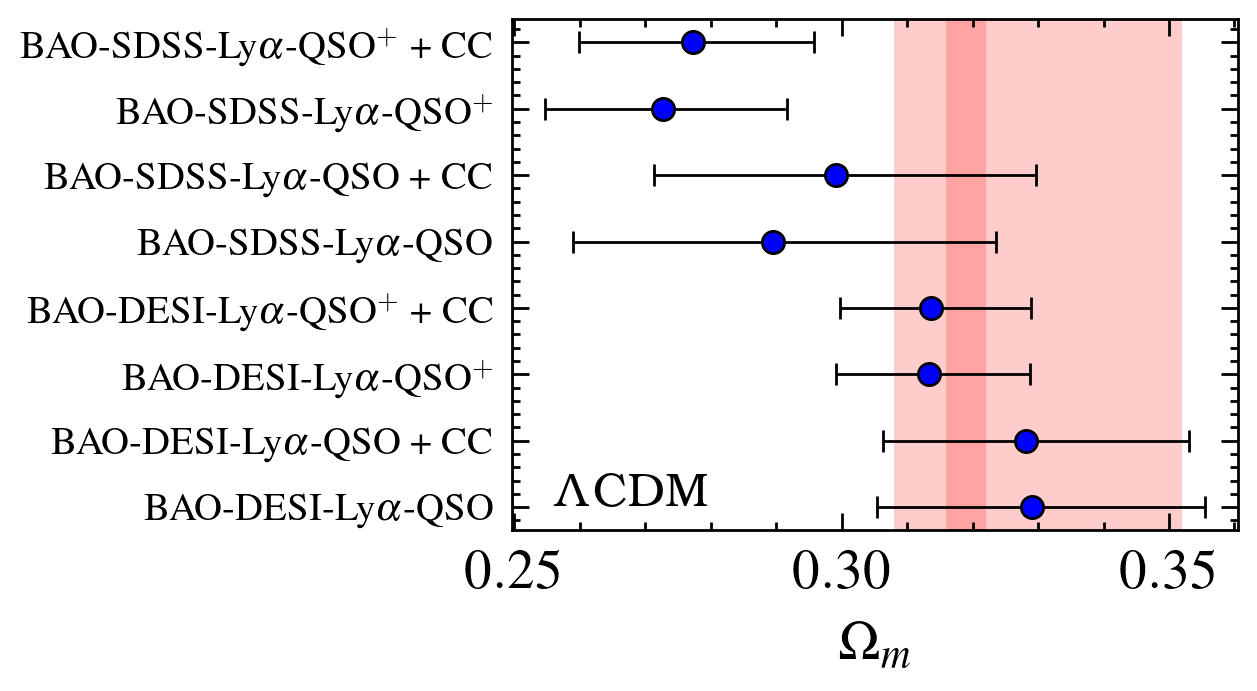}
	\caption{Comparison of matter density estimates for the PEDE and $\Lambda$CDM models using various BAO datasets with and without CC data. The rose bands indicate the Planck 18 ($\Omega_m^{\rm CMB} = 0.315 \pm 0.007$) and SH0ES ($\Omega_m^{\rm SH0ES} = 0.334 \pm 0.018$) estimates. \label{fig:DESISDSSPEDELCDMOm}}
\end{figure*}

\begin{figure}
	\centering
	\includegraphics[width=0.47\textwidth]{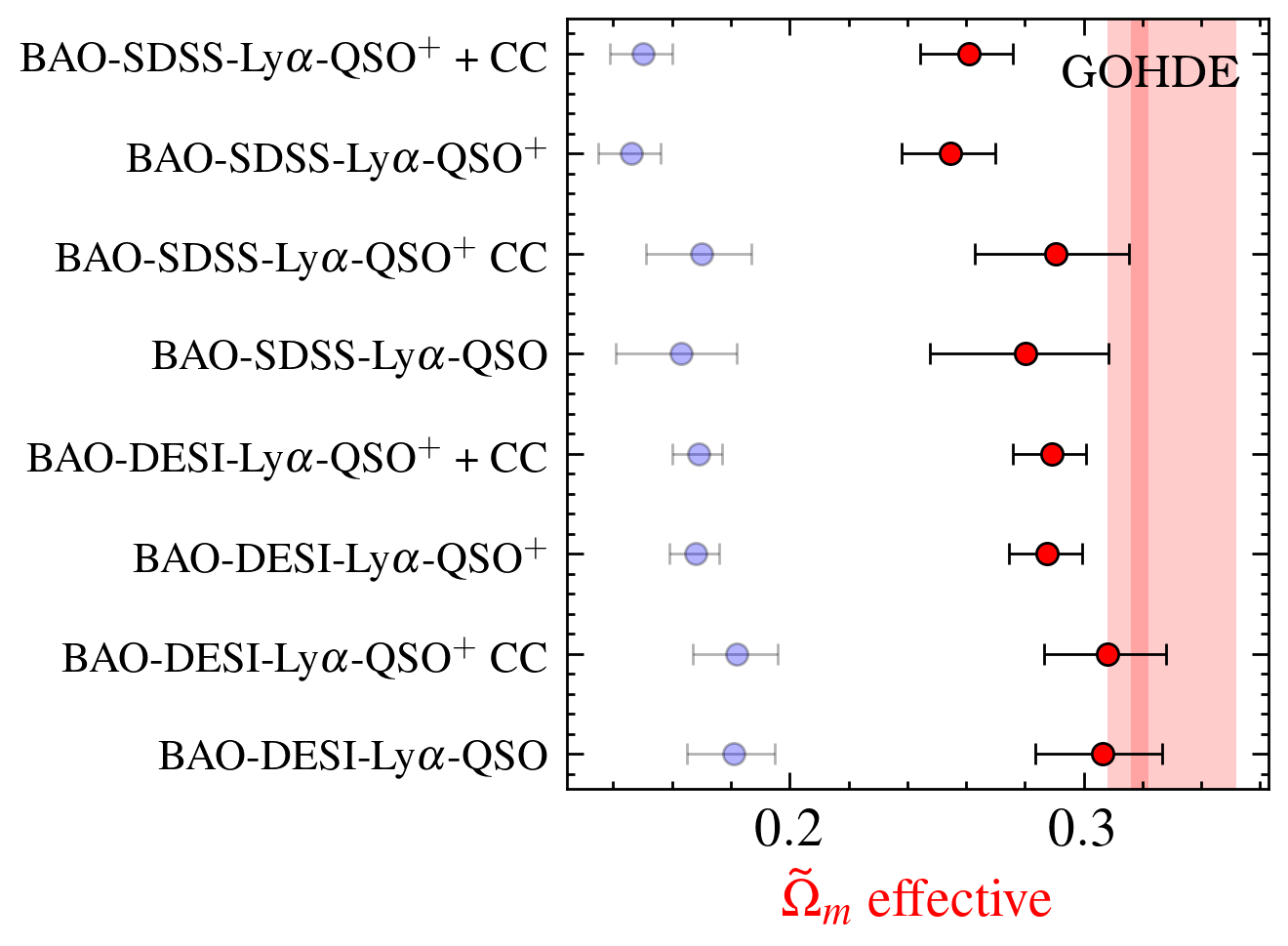}
	\caption{Comparison of matter density estimates for the GOHDE model using various BAO datasets with and without CC data. The light blue markers represent the actual matter density $\Omega_m$, while the red markers denote the effective matter density $\tilde{\Omega}_m$. The rose bands indicate the Planck 18 ($\Omega_m^{\rm CMB} = 0.315 \pm 0.007$) and SH0ES ($\Omega_m^{\rm SH0ES} = 0.334 \pm 0.018$) estimates. \label{fig:DESIGOHDEOm}}
\end{figure}

Regarding $\Omega_m$ estimates, PEDE and GOHDE favor slightly lower values compared to $\Lambda$CDM. The DESI-derived estimates align well with the concordance model, while the SDSS-derived estimates exhibit a relatively larger discrepancy. Once again, the correlation between $H_0$ and $\Omega_m$ is evident. Additionally, these analyses provide mild statistical support for $\Lambda$CDM. The $\Omega_m$ estimates for PEDE and $\Lambda$CDM are shown in FIG. (\ref{fig:DESISDSSPEDELCDMOm}).
In the case of GOHDE, the model distinguishes between $\Omega_m$ (the actual matter density) and $\tilde{\Omega}_m$ (the effective matter density), as depicted in FIG. (\ref{fig:DESIGOHDEOm}). Recognising this distinction is crucial because the CMB shift parameter is directly influenced by the matter density.

Since matter density influences the results, let us now examine the CMB shift parameter, which explicitly depends on it. While the CMB shift parameter alone is insufficient to fully describe the background evolution, as it does not predict $H_0$, it is instrumental in constraining matter density. Specifically, we are interested in the effective matter density.
When fitting models with the CMB shift parameter, we obtain stringent constraints on matter density. Including CC data in the analysis favours a higher value of $H_0$, an effect attributable to the influence of the CMB shift parameter. Previously, analyses using only CC data favoured a higher $\Omega_{m}$ and a lower $H_0$. This reversal highlights the anti-correlation between matter density and $H_0$. Indeed, a model-independent analysis using $H(z)$ data constrains only the combination $\Omega_{m}H_0^2$ \cite{Sahni_2014}. Thus, the observed increase in $H_0$ may not solely reflect the properties of a specific model. This underscores the impact of the correlation between the model's free parameters in shaping the conclusions. The estimates and their corresponding tensions are summarised in TABLE (\ref{tab:CMBCC}).
\begin{table}
	\centering
	\renewcommand{\arraystretch}{1.3}
\begin{ruledtabular}
	\begin{tabular}{llcc}
		Model & Data & $\Omega_m$ or $\tilde{\Omega}_m$ & $H_0$ \\
		\hline
		PEDE   & CC + CMB & $0.261^{+0.008}_{-0.008}$ & $74.165^{+3.585}_{-3.608}$\\
		& CMB      & $0.260^{+0.008}_{-0.008}$ & --\\
		\hline
		GOHDE  & CC + CMB & $0.264^{+0.008}_{-0.006}$ & $73.949^{+3.591}_{-3.571}$ \\
		& CMB      & $0.262^{+0.006}_{-0.006}$ & -- \\
	\end{tabular}
\end{ruledtabular}
\caption{Best fit values of $H_0$ (in km$/$s$/$Mpc) and $\Omega_m$ ($\tilde{\Omega}_m$) derived using the CMB shift parameter alone and in combination with CC data for PEDE and GOHDE models. The table also highlights the tension between these best-fit values and the SH0ES and CMB estimates.\label{tab:CMBCC}}
\end{table}
The CMB shift parameter significantly underestimates the matter density, leading to an overestimation of the Hubble parameter.

Next, we consider the Pantheon$^+$ data sets. Similar to the CMB shift parameter, supernova data alone cannot directly constrain the Hubble parameter. However, due to the strong degeneracy between $H_0$ and the absolute magnitude, supernovae can effectively serve as standardizable candles. Moreover, supernova data can place stringent bounds on the matter density. While there is a negative correlation between matter density and $H_0$, this effect is largely suppressed by the robust nature of supernova constraints.

In previous studies \cite{Li_2019,Li_2020}, particularly with PEDE, a local prior for $H_0$ was often employed, aligning its value closely with the SH0ES estimate. This statistical advantage, combined with the large supernova sample size (nearly 2000 data points), enabled results that mirrored the SH0ES value. However, in this analysis, we refrain from using such a prior. Instead, we combine the Pantheon$^+$ data with various other data sets to explore their combined effects and features. The resulting estimates and statistical metrics are provided in TABLE (\ref{tab:PantheonPfits}).
	\begin{table*}[ht]
		\renewcommand{\arraystretch}{1.3}
	\centering
	\begin{ruledtabular}
		\begin{tabular}{p{1.7cm}cC{2.25cm}cccccccc}
			Data & $\Omega_m$ or $\tilde{\Omega}_m$ & $H_0$ $\left(-M\right)$ & $\sigma_{H_0}^{\rm SH0ES}$ & $\sigma_{H_0}^{\rm CMB}$ & $\sigma_{\Omega_m}^{\rm SH0ES}$ & $\sigma_{\Omega_m}^{\rm CMB}$ & $-2\ln {\cal L}$ & AIC & BIC & DIC \\
			\hline
			\multicolumn{11}{c}{PEDE (Pantheon$^+$ with $\cdots$)} \\
			\hline
			BAO Gal+Ly$\alpha$  & $0.360^{+0.014}_{-0.014}$ & $66.909^{+0.911}_{-0.832}$ $\left(19.460^{+0.025}_{-0.024}\right)$ &  $4.6\sigma$ & $0.5\sigma$ & $1.1\sigma$ & $2.9\sigma$ & $1914.3$ & $1920.3$ & $1940.7$ & $1920.2$ \\
			BAO Gal  & $0.408^{+0.017}_{-0.017}$ & $65.688^{+0.872}_{-0.822}$ $\left(19.483^{+0.024}_{-0.023}\right)$&  $5.6\sigma$ & $1.7\sigma$ & $3.0\sigma$ & $5.1\sigma$ & $1865.2$ & $1871.2$ & $1891.5$ & $1871.3$ \\
			CC  & $0.407^{+0.017}_{-0.017}$ & \cellcolor{red!20!white}$63.841^{+3.222}_{-3.186}$ $\left(19.545^{+0.105}_{-0.109}\right)$&  $2.8\sigma$ & $1.1\sigma$ & $2.9\sigma$ & $5.0\sigma$ & $1771.8$ & $1777.8$ & $1798.1$ & $1777.8$ \\
			CC +~CMB  & $0.297^{+0.008}_{-0.008}$ & \cellcolor{green!20!white}$71.350^{+3.424}_{-3.376}$ $\left(19.344^{+0.101}_{-0.105}\right)$ &  $0.5\sigma$ & $1.2\sigma$ & $1.9\sigma$ & $1.7\sigma$ & $1831.5$ & $1837.5$ & $1857.7$ & $1837.4$ \\
			\hline
			\multicolumn{11}{c}{GOHDE (Pantheon$^+$ with $\cdots$)} \\
			\hline
			BAO Gal+Ly$\alpha$ & $0.354^{+0.016}_{-0.015}$ & $67.883^{+0.885}_{-0.895}$ $\left(19.440^{+0.025}_{-0.025}\right)$ & $4.0\sigma$ & $0.5\sigma$ & $0.8\sigma$ & $2.2\sigma$ & $1930.7$ & $1936.7$ & $1957.1$ & $1936.7$ \\
			BAO Gal  & $0.431^{+0.020}_{-0.018}$ & $65.753^{+0.880}_{-0.870}$ $\left(19.487^{+0.025}_{-0.024}\right)$ &  $5.5\sigma$ & $1.6\sigma$ & $3.7\sigma$ & $5.6\sigma$ & $1870.5$ & $1876.5$ & $1896.8$ & $1876.5$ \\
			CC  & $0.430^{+0.020}_{-0.018}$ & \cellcolor{red!20!white}$63.127^{+3.194}_{-3.283}$ $\left(19.577^{+0.107}_{-0.114}\right)$ &  $3.0\sigma$ & $1.3\sigma$ & $3.6\sigma$ & $5.5\sigma$ & $1777.4$ & $1783.4$ & $1803.7$ & $1783.5$ \\
			CC +~CMB  & $0.286^{+0.007}_{-0.007}$ & \cellcolor{green!20!white} $72.278^{+3.295}_{-3.393}$ $\left(19.323^{+0.097}_{-0.105}\right)$&  $0.3\sigma$ & $1.4\sigma$ & $2.5\sigma$ & $2.8\sigma$ & $1838.8$ & $1844.8$ & $1865.0$ & $1844.6$ \\
		\end{tabular}
	\end{ruledtabular}
\caption{Best-fit values of $H_0$ (in km$/$s$/$Mpc) and $\Omega_m$ ($\tilde{\Omega}_m$) for PEDE and GOHDE models obtained using Pantheon$^+$ data combined with other data sets. The table also includes the respective $-2\ln {\cal L}$, AIC, BIC, and DIC values, highlighting the tension between the best-fit results and the SH0ES and CMB estimates of these parameters.\label{tab:PantheonPfits}}
\end{table*}

In contrast to the earlier scenario, where the combination of BAO Gal and Ly$\alpha$ data favoured a higher $H_0$ value, the Hubble tension re-emerges when the Pantheon$^+$ sample is included. While BAO Ly$\alpha$ continues to mitigate the tension, BAO Gal alone produces estimates less favourable than the Planck results. The influence on $\Omega_{m}$ estimates is significant in both cases, highlighting the statistical importance of the data. The Pantheon$^+$ sample imposes a stringent constraint on matter density, and its inherent degeneracy between $H_0$ and $M$ impacts the models' ability to effectively resolve the Hubble tension.

Using Pantheon$^+$ data alongside any $H(z)$ measurements constrains the $\Omega_{m}H_0^2$ parameter, emphasising the need for strong, independent constraints on $\Omega_{m}$ to further break the degeneracy. Incorporating the CMB shift parameter resolves the Hubble tension without introducing substantial tension in the matter density estimates, as evidenced in the case of CC + CMB data shown in TABLE (\ref{tab:PantheonPfits}). Between the two models, PEDE is statistically favoured over GOHDE.

\subsection{Minimal Extensions}
\begin{figure}
	\centering
\includegraphics[width=0.23\textwidth]{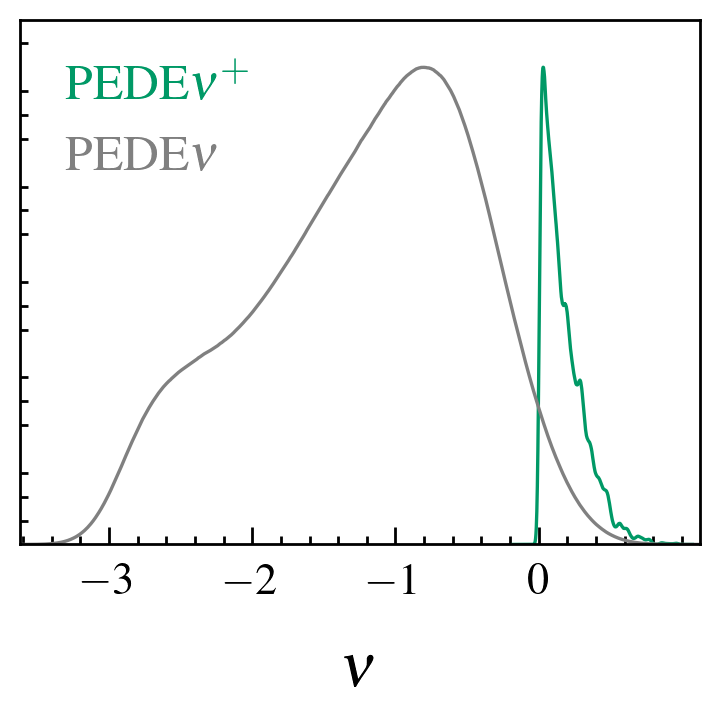}
\includegraphics[width=0.23\textwidth]{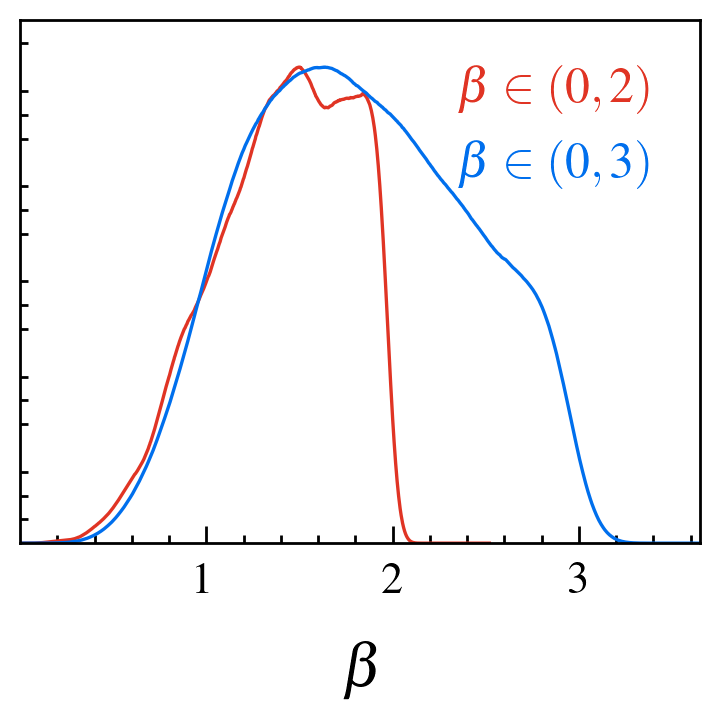}
\caption{Posterior distributions of the additional free parameters $\nu$ (PEDE) and $\beta$ (GOHDE) using the full OHD dataset, analysed under various uniform priors. \label{fig:ExtraFreeParams}}
\end{figure}

When considering the complete framework, both PEDE and GOHDE introduce one additional free parameter. In the previous analysis, these parameters were held fixed when addressing the Hubble tension. Here, we examine the broader perspective by allowing these parameters to vary and analysing their impact on the conclusions. To simplify the analysis, we focus on the OHD dataset, which combines CC, BAO Gal, and BAO Ly$\alpha$ data. Allowing the free parameters of PEDE and GOHDE to vary, we fit the models using the OHD dataset. The posterior distributions of these free parameters are presented in FIG. (\ref{fig:ExtraFreeParams}). Across different prior settings, the free parameter of PEDE ($\nu$) tends to favour negative values, while the free parameter of GOHDE ($\beta$) is inclined toward values exceeding $2/3$.

For the PEDE model, when the prior on $\nu$ is restricted to positive values, the best-fit parameters are $H_0 \sim 70.9$ km$/$s$/$Mpc and $\Omega_{m} \sim 0.26$. However, $\nu$ itself remains poorly constrained, with a maximum likelihood value of $\nu \sim 0.0021$. This effectively reduces the model to $\Lambda$CDM, as $\nu \rightarrow 0$.
Expanding the prior to include negative values for $\nu$, the best-fit shifts to $\nu \sim -1.12$, with corresponding values of $H_0 \sim 62.95$ km$/$s$/$Mpc and $\Omega_{m} \sim 0.31$. In this case, the model fails to resolve the Hubble tension when $\nu$ is left entirely free. The interplay of a negative correlation between $\nu$ and $\Omega_{m}$ and a positive correlation between $\nu$ and $H_0$ significantly impacts the best-fit values. Consequently, in its original formulation, the PEDE model does not effectively address the Hubble tension, although it maintains consistency in matter density estimates. This is yet another manifestation of the correlation between the free parameters.
\begin{figure*}
	\centering
	\includegraphics[width=0.47\textwidth]{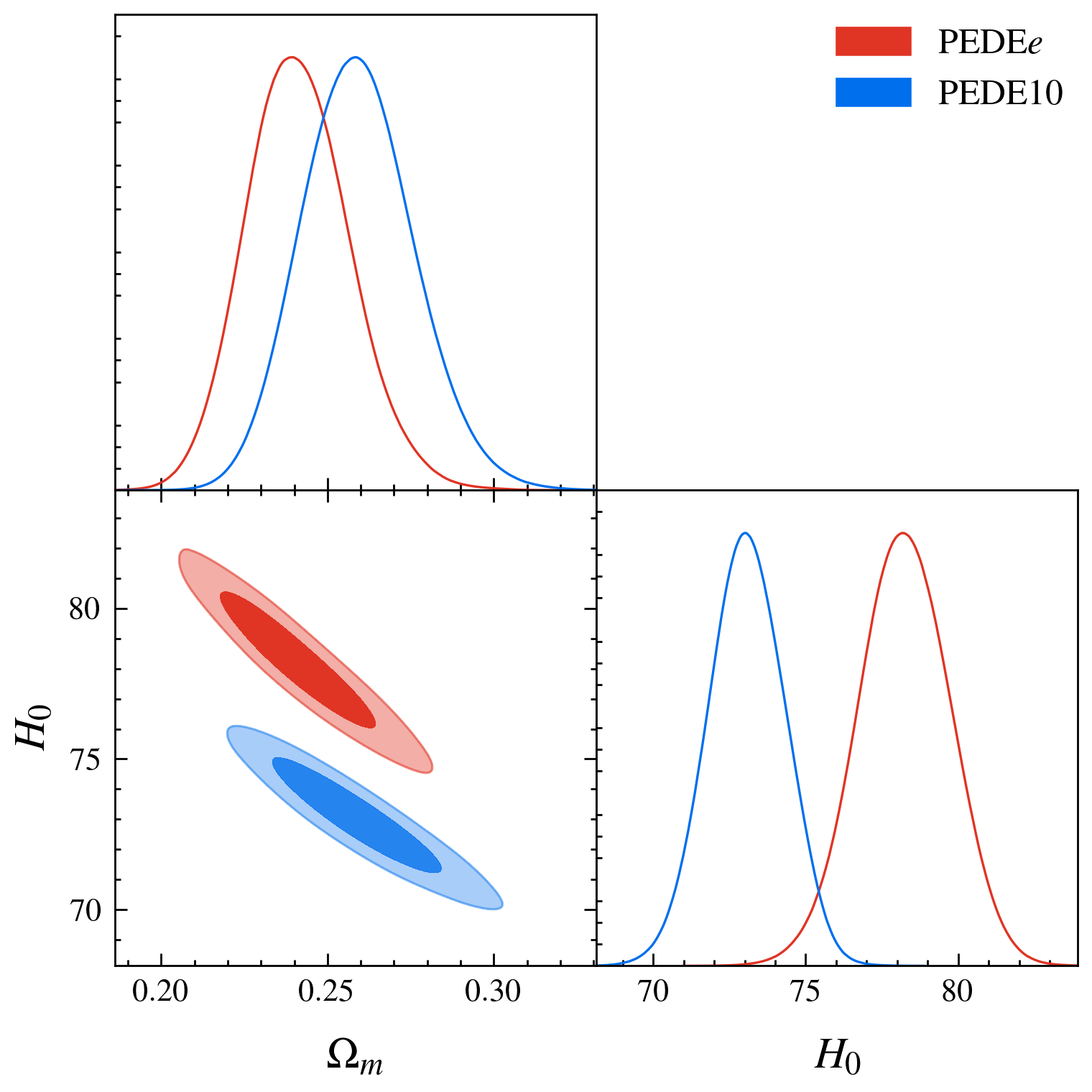}
	\includegraphics[width=0.47\textwidth]{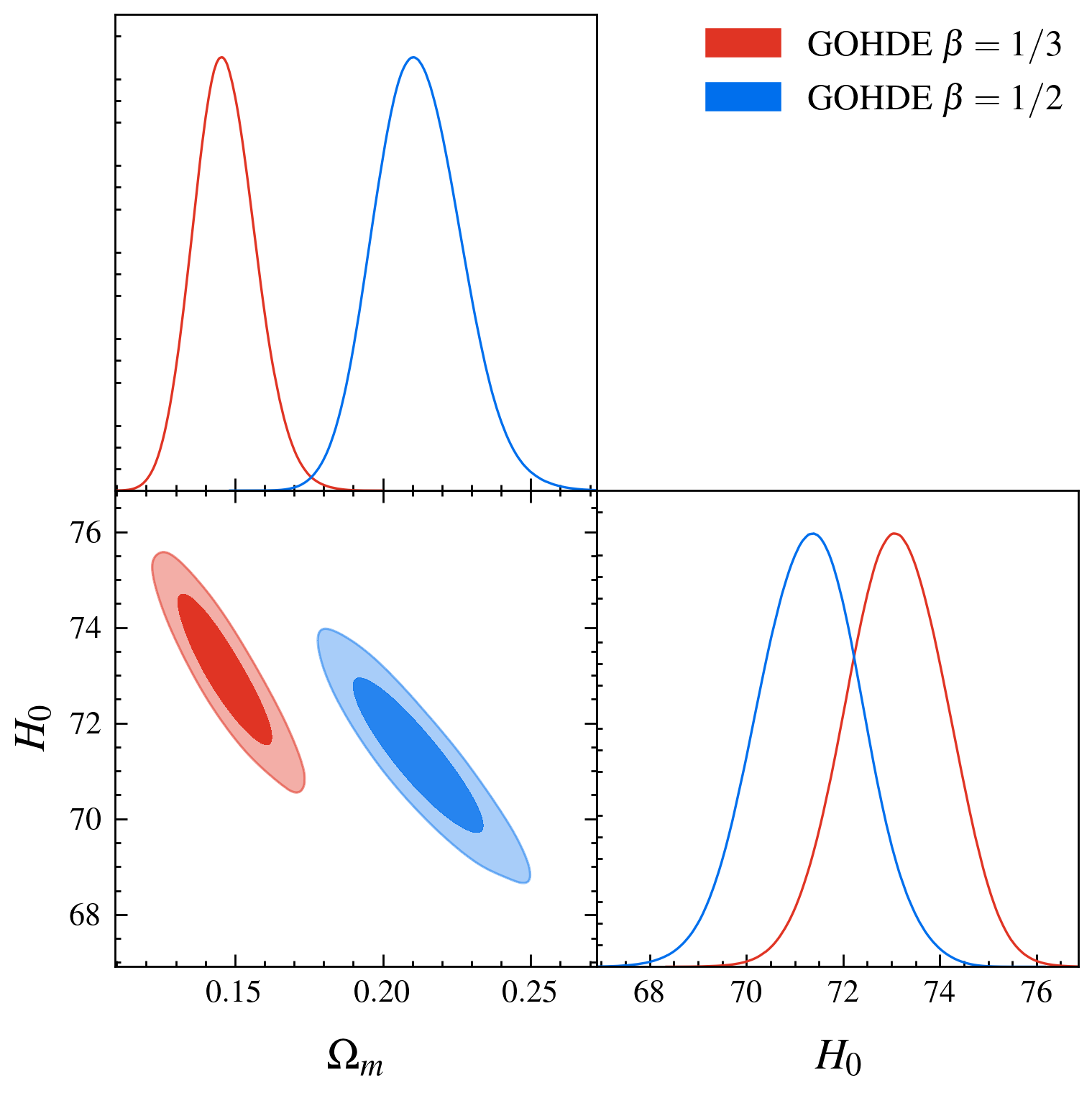}
	\caption{Corner plots of $\Omega_{m}$ and $H_0$ for fixed values of $\nu$ and $\beta$ in the generalized PEDE and GOHDE models, respectively. Note that for the GOHDE model, the plots show $\Omega_{m}$ rather than the effective matter density $\tilde{\Omega}_m$. \label{fig:effectofvb}}
\end{figure*}

The behaviour of GOHDE closely mirrors that of PEDE. When $\beta$ is treated as a free parameter, the best-fit values are $\beta \sim 1.45$, $H_0 \sim 64.9$ km$/$s$/$Mpc, and $\Omega_{m} \sim 0.45$. This outcome reflects the influence of parameter correlations: there is a positive correlation between $\beta$ and the matter density, while $\beta$ and $H_0$ exhibit a negative correlation. Consequently, the model predicts a higher matter density ($\Omega_{m}$) and a lower Hubble parameter ($H_0$). Extending the prior to higher values, the best-fit shifts to $\beta \sim 1.79$, $H_0 \sim 63.7$ km$/$s$/$Mpc, and $\Omega_{m} \sim 0.5$. Clearly, like PEDE, GOHDE does not naturally resolve the Hubble tension. Particularly when the additional free parameter is assigned a broad uniform prior.

However, for both cases, the effective matter density remains close to $\tilde{\Omega}_m \sim 0.27$, computed using the expression $\tilde{\Omega}_m = {2\Omega_m}/({3\beta - 3\beta\Omega_m + 2\Omega_m})$, as outlined in earlier sections. The posterior distributions shown in FIG. (\ref{fig:ExtraFreeParams}) highlight how strongly the results depend on the chosen prior bounds for $\beta$. Furthermore, the parameter value $\beta = 2/3$ holds particular significance, as GOHDE reduces to the $\Lambda$CDM model for this value. Any $\beta > 2/3$ results in a lower $H_0$. Similarly, for PEDE, $\nu = 0$ recovers the $\Lambda$CDM model, with lower $\nu$ values yielding smaller $H_0$ estimates. Thus, the choice of prior of $\nu$ and $\beta$ significantly influences the results.

To illustrate this, consider different parameter values. For instance, setting $\beta = 1/2 < 2/3$ or $\nu = 1 > 0$ results in $H_0$ estimates higher than those predicted by $\Lambda$CDM. This shift is visualised in FIG. (\ref{fig:effectofvb}), which shows $H_0$ estimates for various $\nu$ and $\beta$ values. For PEDE with $\nu = 1/\ln e$, the $H_0$ value is significantly higher than for PEDE with $\nu = 1/\ln 10$, the case discussed earlier. Similarly, for $\beta = 1/2$, $H_0$ is higher compared to $\Lambda$CDM, with an even greater increase observed for $\beta = 1/3$.

Choosing the specific values $\beta=1/3$ and $\nu=1/\ln 10$ was critical in the earlier data analysis. A dataset that originally suggests a lower value of $H_0$ will predict a higher value when these parameters are fixed accordingly. For example, when $\nu = 1/\ln e = 1$, we obtain $H_0 = 78.23 \pm 1.5$ km/s/Mpc and $\Omega_m = 0.24 \pm 0.02$. Similarly, for the GOHDE model with $\beta = 1/2$, the estimates are $H_0 \sim 71.3$ km/s/Mpc and $\Omega_m \sim 0.26$, and so forth.

\begin{figure}
	\centering
	\includegraphics[width=0.47\textwidth]{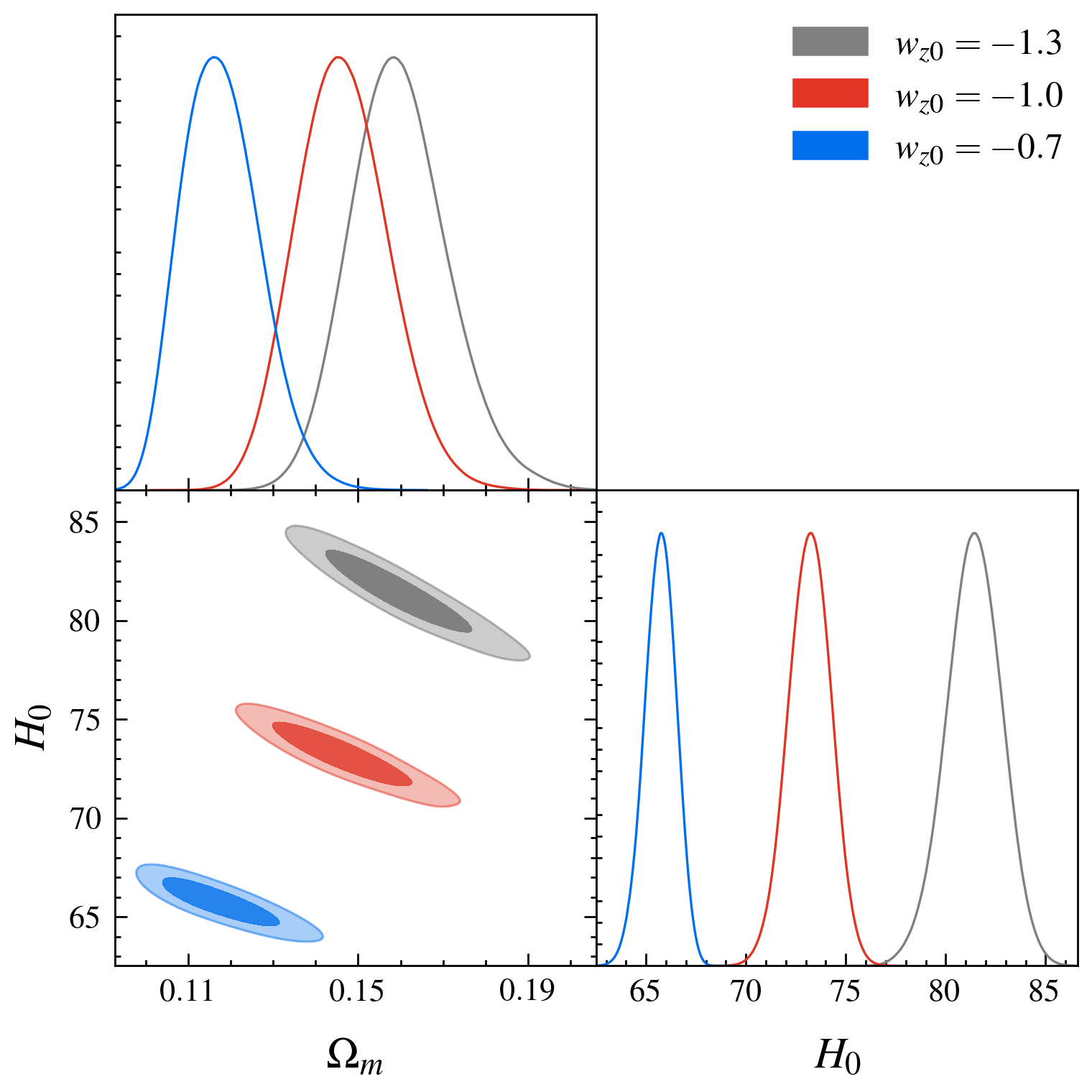}
	\caption{Impact of varying the parameter $w_{z0}$ with fixed $\beta=1/3$ on the GOHDE model fits using the OHD data set. \label{fig:effectofwz0}}
\end{figure}
In the full GOHDE model, there is an additional free parameter, $w_{z0}$, representing the present value of the dark energy equation of state parameter. For simplicity, we fixed $w_{z0} = -1$, but variations in this parameter can also significantly affect parameter estimates. This effect is illustrated in FIG. (\ref{fig:effectofwz0}). Clearly, a more phantom-like present value of the dark energy equation of state ($w_{z0} < -1$) leads to a higher estimated $H_0$. For instance, with $w_{z0} = -1.3$, the Hubble parameter can reach values as high as $H_0 \sim 81$ km$/$s$/$Mpc. Therefore, fixing these originally free parameters to specific values has significantly influenced the conclusion that these models could settle the Hubble tension.

To put this into perspective, consider the $w$CDM model. When $w = -1$, it corresponds to the $\Lambda$CDM model, which predicts an $H_0$ of about 67 km$/$s$/$Mpc. Lowering $w$ to -1.3 increases the predicted $H_0$ to nearly 76 km$/$s$/$Mpc. Although model selection typically depends on various statistical criteria, current data show only a marginal preference between these models. Therefore, more robust data or a more comprehensive analysis is needed to draw definitive conclusions. One important dataset not yet included is the full CMB power spectrum. While a detailed analysis is beyond the scope of this work, it sheds light on why these phenomenological models may have an edge over simpler $w$CDM models.

\begin{figure}[h]
	\centering
	\includegraphics[width=0.47\textwidth]{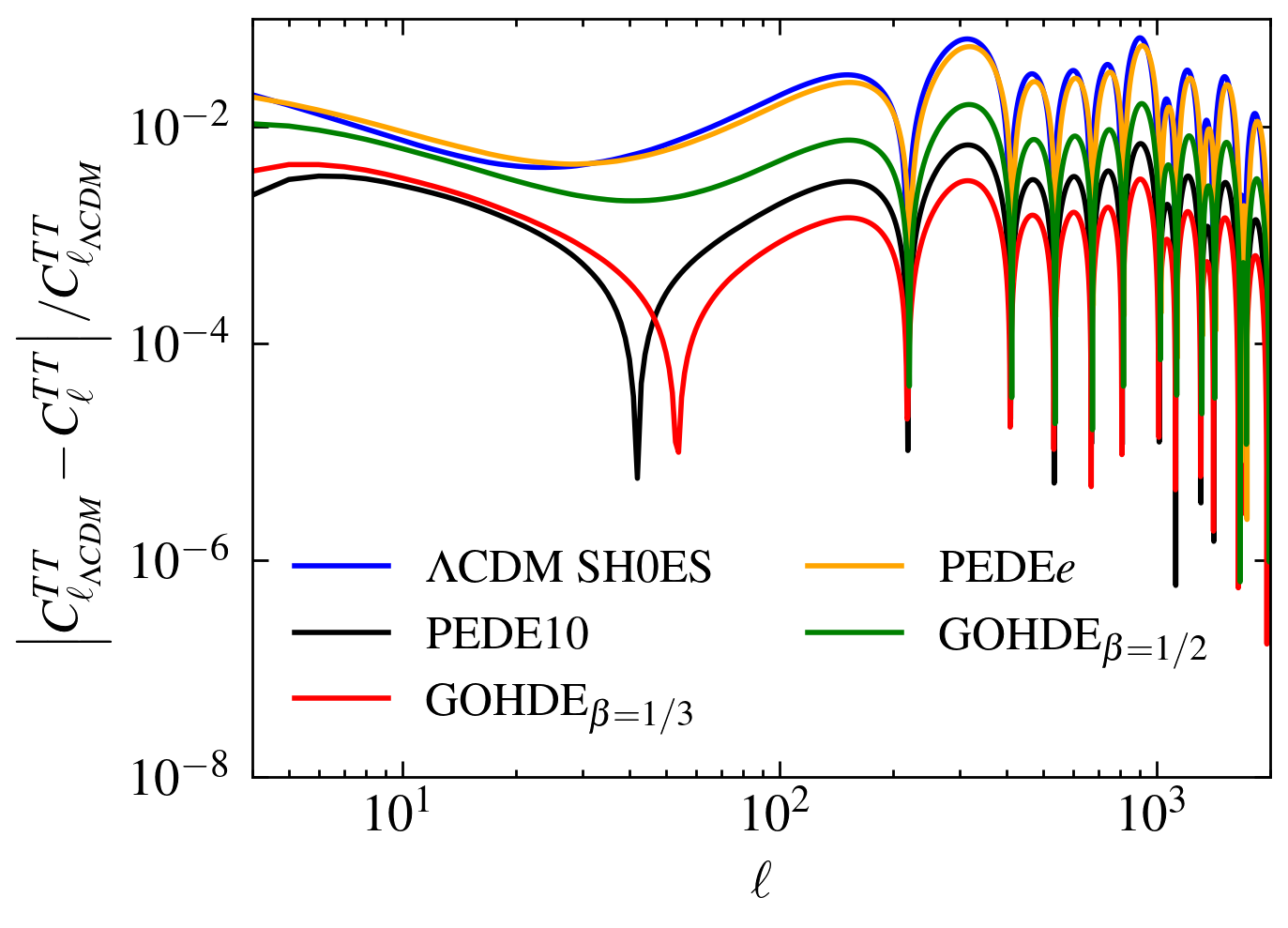}
	\caption{Deviation of the CMB TT power spectrum from the Planck 2018 $\Lambda$CDM best-fit for various PEDE and GOHDE model variants, alongside the $\Lambda$CDM model with the SH0ES prior.\label{fig:CMBsample}}
\end{figure}

The simplest approach is to modify CAMB \cite{CAMBPython}. For illustration, let us consider the Hubble parameter fixed at 73 km$/$s$/$Mpc. We then ask: how much deviation from the standard $\Lambda$CDM do we expect when enforcing this SH0ES estimate under various models, while keeping all other cosmological parameters unchanged? To implement this, we use the \href{https://raw.githubusercontent.com/cmbant/CAMB/master/inifiles/planck_2018.ini}{Planck 18} best fit configuration file as the reference $\Lambda$CDM and modify it accordingly to incorporate our GOHDE and PEDE models. The varying dark energy equation of state is implemented using the \texttt{DarkEnergyPPF} function from the \texttt{camb.dark$_-$energy} module, defined as a function of the scale factor. The PPF (parametrised post-Friedmann dark energy) framework allows phantom transitions without causing instabilities \cite{PhysRevD.78.087303}. We consider two variants, each of PEDE and GOHDE, alongside $\Lambda$CDM to demonstrate that these alternatives provide a better fit for a higher $H_0$. As shown in FIG. (\ref{fig:CMBsample}), both PEDE${10}$ and GOHDE${\beta=1/3}$ perform significantly better in achieving a larger $H_0$ without deviating substantially from the reference $\Lambda$CDM.

\subsection{Interaction picture}
A common consensus is that some degree of phantom behaviour in dark energy is necessary to resolve the Hubble tension, with the amount of phantomness roughly correlating with how much we want to increase the value of $H_0$. However, there is still no universal agreement on whether dark energy must be fully phantom. In the PEDE model, we have an emergent phantom dark energy component that remains phantom and asymptotically approaches a de Sitter phase. In contrast, the GOHDE model features a dynamical dark energy that tracks the total energy density, behaving like a quintessence field in the past and crossing into the phantom regime in the future by design. Beyond this phantom characteristic, there is no single generic property of these models that fully explains their ability to raise $H_0$. Given that these models exhibit a background-level degeneracy with interacting dark sector models, it is important to study them from that perspective as well. In this section, we reframe the previous PEDE and GOHDE models as interacting dark energy scenarios and explore their implications.

Interestingly, these models exhibit a distinctive behaviour when we consider the interaction term $\tilde{Q} = 3H \gamma_{\Lambda} \tilde{\rho}_m$ introduced earlier. A positive $\tilde{Q}$ corresponds to energy transfer from dark energy to dark matter, while a negative value indicates the opposite flow. In this expression, there are three components: $H$, $\tilde{\rho}_m$, and $\gamma_{\Lambda}$. Since both $H$ and $\tilde{\rho}_m$ are inherently positive, the sign of the interaction term is determined solely by the interaction strength $\gamma_{\Lambda}$. When $\gamma_{\Lambda} = 0$, the models reduce to the standard $\Lambda$CDM scenario, providing a unified framework. Exploring the parameter space that can alleviate the Hubble tension reveals a sudden shift in the interaction behaviour.
\begin{figure}[b]
	\centering
	\includegraphics[width=0.47\textwidth]{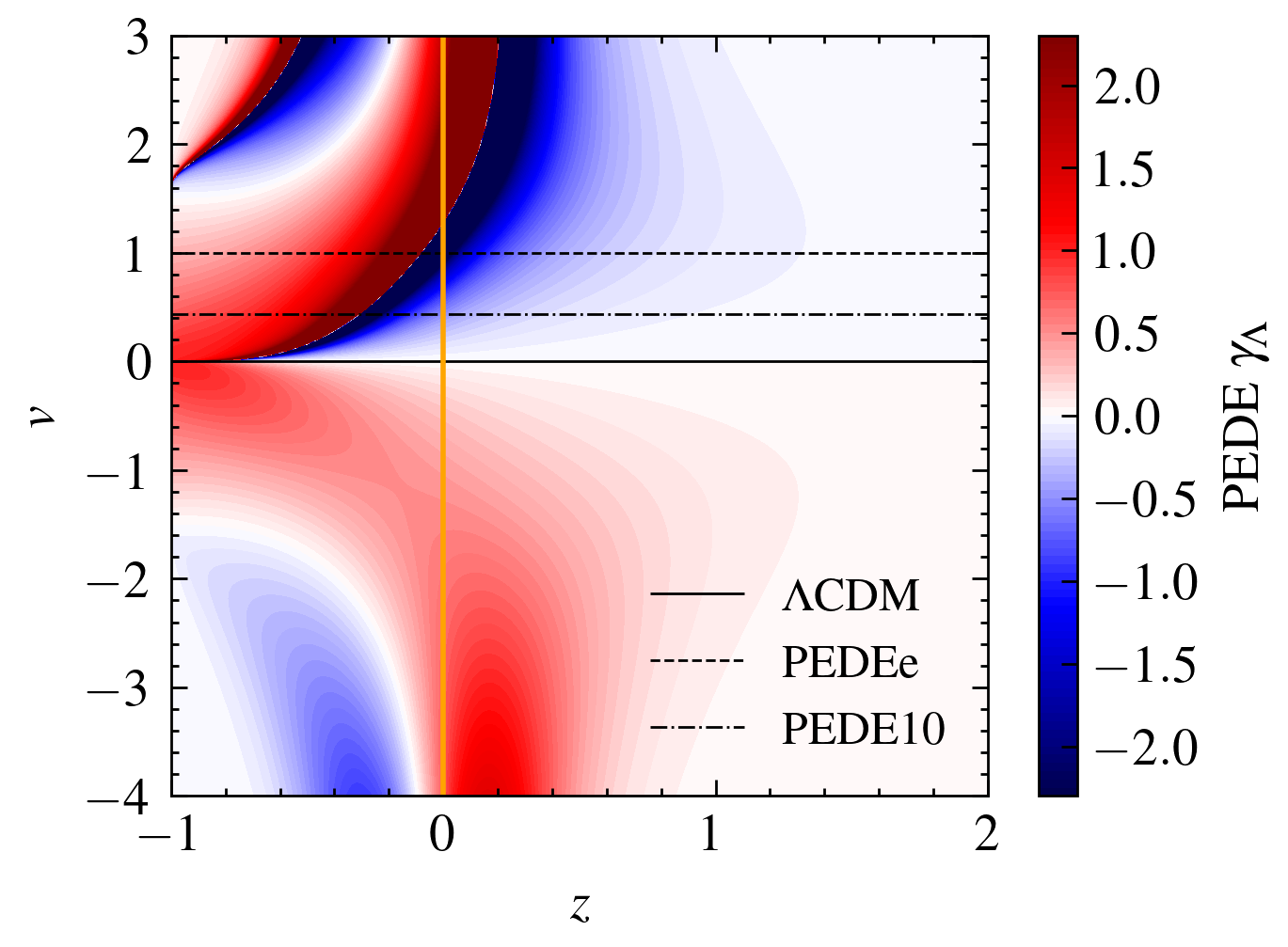}
	\caption{Interaction strengths $\gamma_{\Lambda}$ plotted as functions of the free parameter $\nu$ and redshift $z$ for the generalized PEDE model, assuming $\Omega_{m} = 0.3$. \label{fig:IntglPEDE}}
\end{figure}
Consider the plots in FIG. (\ref{fig:IntglPEDE}). For the PEDE model, $\Lambda$CDM appears as the special case when $\nu = 0$, corresponding to zero interaction. As $\nu$ increases, we observe that in the past, $\gamma_{\Lambda}$ was negative, diverging to negative infinity before abruptly switching to a positive value, indicating the presence of a pole in the expression for $\gamma_{\Lambda}$. 
\begin{figure}
	\centering
	\includegraphics[width=0.47\textwidth]{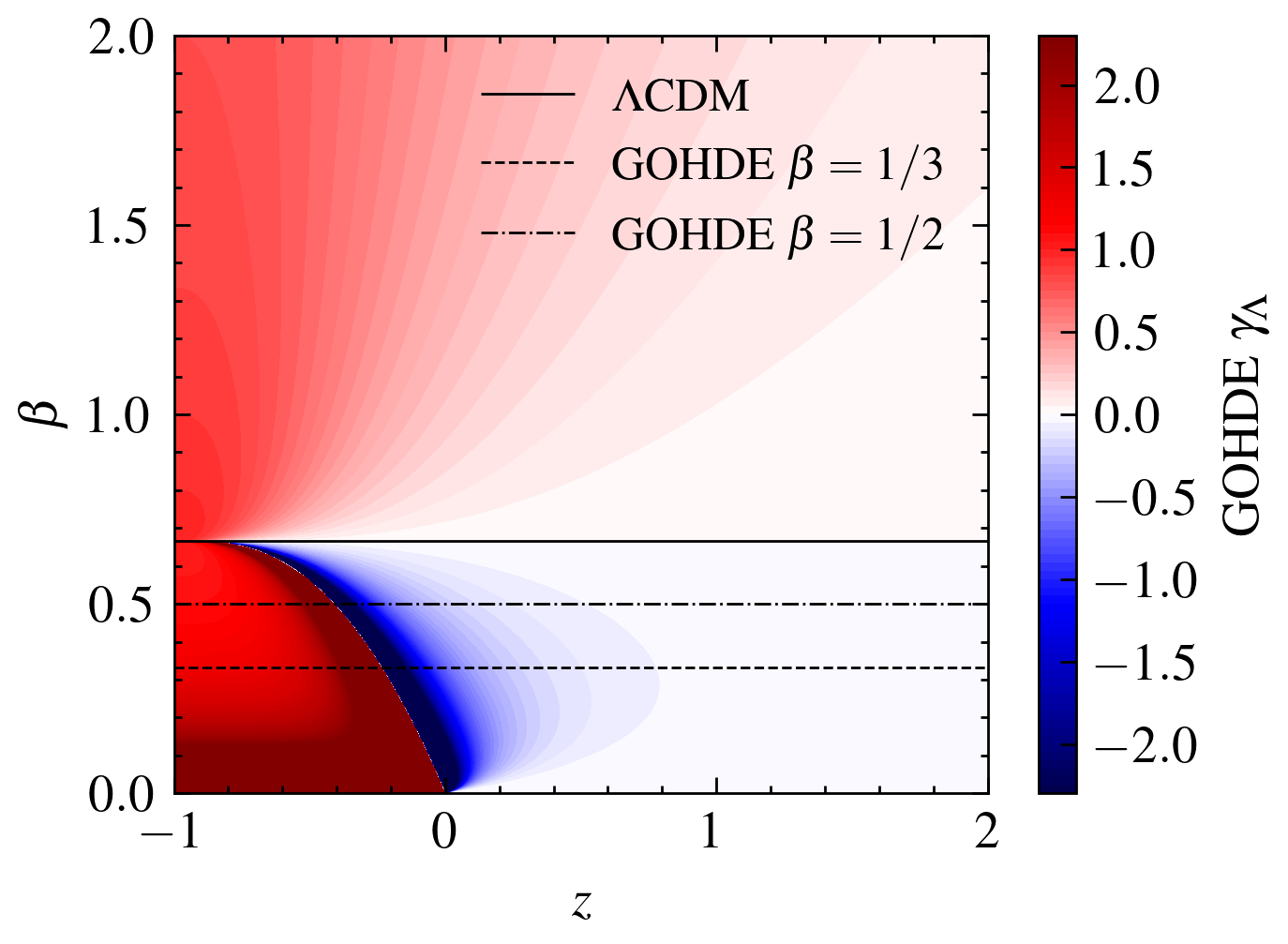}
	\caption{Interaction strength $\gamma_{\Lambda}$ as a function of free parameter $\beta$ and $z$ for GOHDE model, with $\Omega_{m}=0.15$. This particular value of $\Omega_{m}$ corresponds to an effective matter density $\tilde{\Omega}_m=0.26$. \label{fig:IntglGOHDE}}
\end{figure}
Similarly, for the GOHDE model, FIG. (\ref{fig:IntglGOHDE}), when $\beta = 2/3$, the interaction vanishes, but lowering $\beta$ causes a sudden transition from negative to positive interaction.
\begin{figure}[b]
	\centering
	\includegraphics[width=0.47\textwidth]{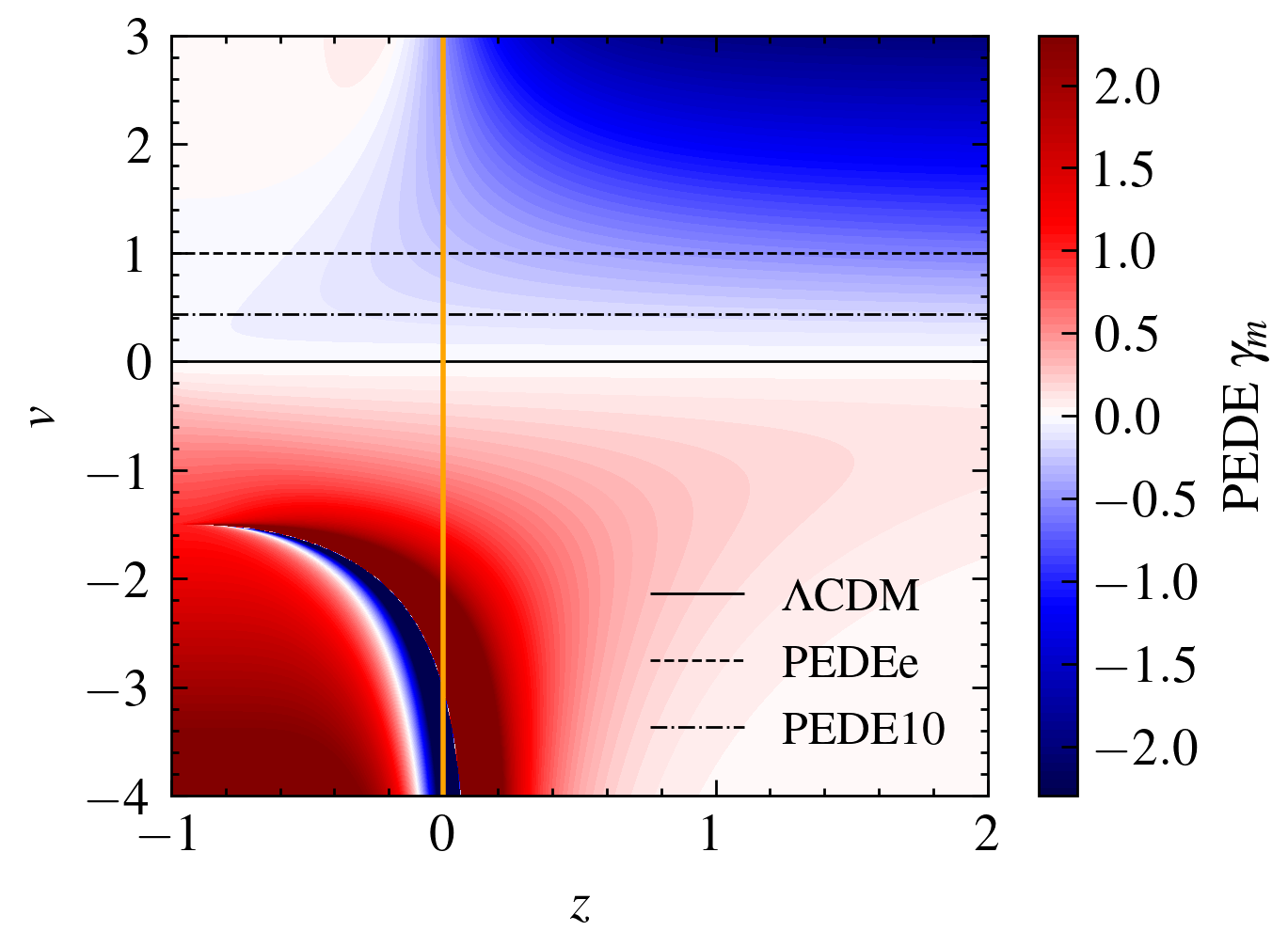}
	\caption{Interaction strengths $\gamma_{m}$ plotted as functions of the free parameter $\nu$ and redshift $z$ for the generalized PEDE model, assuming $\Omega_{m} = 0.3$.\label{fig:gmpede}}
\end{figure}
This suggests that a dynamic interaction of the form $\tilde{Q} = 3H \gamma_{\Lambda} \tilde{\rho}_m$, where $\gamma_{\Lambda}$ depends on redshift and can change sign with a pole, can produce the desired values of the Hubble parameter. Conversely, when the free parameters vary in the opposite direction, no such poles occur. The number of times this sign change can happen depends on the specific model, since $\gamma_{\Lambda}$ is a function of the model’s free parameters. This behaviour does not apply to $\gamma_{m}$, which can either be constant, as in the GOHDE model, or independent of $\Omega_m$, as in the PEDE model.

What we find particularly intriguing are these sudden jumps in the interaction, which signify a reversal of energy flow between dark energy and dark matter. These transitions carry additional implications in a broader context. Since $\tilde{\rho}_m$ remains positive by definition, and $\gamma_{\Lambda} = \tilde{f} \tilde{\rho}_{\Lambda} / (\tilde{\rho}_m + \tilde{\rho}_{\Lambda})$, where $\tilde{\rho}_m + \tilde{\rho}_{\Lambda}$ represents the total energy density (which is conserved and thus never negative), for $\gamma_{\Lambda}$ to change sign, the product $\tilde{f} \tilde{\rho}_{\Lambda}$ must become negative. Given that specific energy densities are typically non-negative throughout cosmic evolution, such a jump is non-trivial. For this to occur, $\tilde{f}$ must undergo a sign change. According to the definition, this implies,
\begin{equation}
\tilde{f}\begin{cases}
<0, \implies \frac{a\tilde{r}'}{3\tilde{r}}<-1\\
>0, \implies \frac{a\tilde{r}'}{3\tilde{r}}>-1\\
\end{cases}
\end{equation}
Thus, as a function of the scale factor, the sign flip of $\tilde{f}$ occurs when $\frac{\tilde{r}'}{\tilde{r}} = -\frac{3}{a}$. Interestingly, this can be rewritten as,
\begin{align}
\frac{\tilde{r}'}{\tilde{r}}=\frac{d}{da}\left(\ln \tilde{r} \right)=-\frac{3}{a}\\
\ln \tilde{r} = -3\ln a + \ln \mathcal{C}\implies
\tilde{r} = \frac{\mathcal{C}}{a^3}
\end{align}
From the definition of $\tilde{r}$, we conclude that the sign flip occurs when the ratio of matter density to dark energy density deviates from the $\Lambda$CDM behaviour. These transitions take place in the future if we want the Hubble parameter to approach the higher values reported by SH0ES. The diagram shows that if this transition does not occur, the model remains in a $\Lambda$CDM-like regime. As we shift this transition closer to the present epoch ($z \to 0$) from the future, higher values of $H_0$ are obtained. In summary, within the interaction framework, the model exhibits negative interaction in the past, followed by a positive interaction in the future.

Does this abrupt sign change signify something deeper? Yes -- it corresponds to the maximisation of horizon entropy and marks the onset of the phantom phase. The Hubble horizon entropy is defined as a quantity proportional to the area of the Hubble horizon. In the $\Lambda$CDM model, this entropy reaches its maximum only in the de Sitter limit, while models with ever-increasing horizon area correspond to quintessence behaviour. When phantom fields dominate, the second law of thermodynamics is violated, and entropy maximisation no longer holds.

For the GOHDE model with $\beta=1/3$, which predicts a phantom future, the entropy decreases over time. In contrast, PEDE models with an asymptotic de Sitter phase exhibit entropy that initially increases, reaches a maximum, and then settles to a finite value. Depending on the free parameter, the entropy can show multiple extrema. Importantly, this entropy maximisation is captured by the interaction strength $\gamma_{\Lambda}$, and the timing of this maximum can affect the best-fit value of $H_0$.

In summary, whether the phantom dark energy phase occurs in the past, present, or future is not the key factor; rather, it is the violation of the second law of thermodynamics that guarantees relaxation of the Hubble tension. Notably, a violation of the second law in the past corresponds to a significantly higher value of $H_0$. 

\begin{figure}[h]
	\centering
	\includegraphics[width=0.47\textwidth]{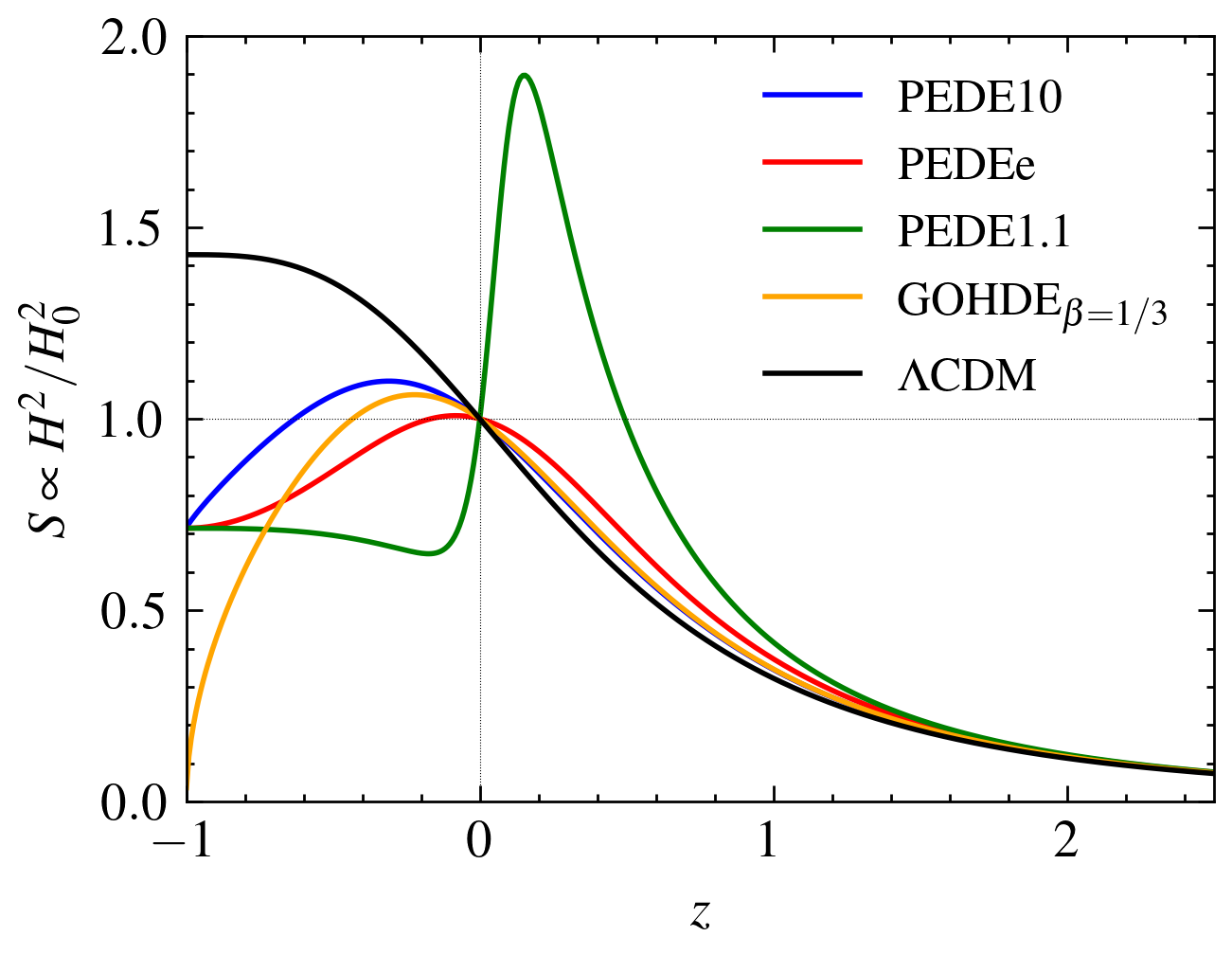}
	\caption{Evolution of the horizon entropy normalised to unity at present for PEDE models for different $\nu$ values, GOHDE with $\beta=1/3$, and $\Lambda$CDM models. \label{fig:entropy}}
\end{figure}

In FIG. (\ref{fig:entropy}), we observe that for the GOHDE and PEDE models, the horizon entropy does not reach its maximum value at the asymptotic future, unlike in the $\Lambda$CDM model, where the maximum entropy is attained in the final de Sitter phase, consistent with the second law of thermodynamics. For PEDE, depending on the free parameter $\nu$, the entropy rises to a maximum and then settles to a lower finite value, while for GOHDE, it first increases, attains its maximum, and then drops over time. This implies that both models, which help relax the Hubble tension, ultimately violate the second law of thermodynamics in the future.

Therefore, having a past or future phantom dark energy equation of state is not an essential requirement. Instead, a future violation of the second law appears inevitable if these phenomenological models are chosen to address the Hubble tension at the background level. Conversely, an earlier (past) violation would push the Hubble parameter in the opposite direction, worsening the tension. It is important to note that this violation is limited to the horizon entropy; incorporating additional entropy production mechanisms may preserve the overall thermodynamic consistency -- a topic for future inquiry.

Interestingly, the interaction framework captures this entropy maximisation, or the violation of the second law, which is not widely discussed in the literature. Thus, the divergence of the interaction coefficient $\gamma_{\Lambda}$ should not be interpreted as a complete energy transfer between dark energy and dark matter, but rather as the onset of a change in the rate of horizon entropy evolution. This transition can also be seen as the onset of a violation of the null energy condition (NEC). Consequently, the fundamental aspect for resolving the Hubble tension is not simply the presence of phantom dark energy, but ``when'' the NEC is violated. A past violation leads to an overestimation of $H_0$, while a future violation sets $H_0$ to a value more consistent with observational constraints. The $\Lambda$CDM model, which has no such violation, simply fails to resolve the tension.

These subtleties are not captured by the simpler interaction parametrisation $Q=3H\gamma_m \rho_{\Lambda}$. For instance, in GOHDE, $\gamma_m$ is constant, and in PEDE, $\gamma_m$ shows no distinctive behaviour that singles out models capable of alleviating the Hubble tension (see FIG. \ref{fig:gmpede}).

In summary, the violation of the second law of horizon thermodynamics is equivalent to violating the null energy condition, which requires $\rho + p \geq 0$. The NEC implies that the derivative of the Hubble parameter should be negative, i.e., the Hubble parameter must decrease over time. A violation of NEC corresponds to an increasing Hubble parameter. This particular phantom behaviour is the key feature needed for models like PEDE and GOHDE to raise $H_0$ and potentially resolve the Hubble tension.

\section{Conclusions}

We began by introducing the Hubble tension and exploring minimal extensions to the $\Lambda$CDM model that can predict a higher value of $H_0$, specifically the PEDE and GOHDE models. Our goal was to evaluate their performance across various data sets and assess whether these models yield consistent $H_0$ estimates independent of the data choice.

We found that cosmic chronometer data favours a best-fit $H_0$ close to the Planck 18 CMB estimate but with larger uncertainties, effectively masking any tension due to the wide error bars. However, when using BAO galaxy clustering data, the best-fit $H_0$ remains near the Planck 18 value, which creates an apparent tension with the higher SH0ES measurement. Both the PEDE and GOHDE models predict values of $H_0$ larger than $\Lambda$CDM but still fall short of fully resolving the tension.

Introducing BAO Ly$\alpha$ measurements at redshift $\sim 2.3$ changes the picture: by adding these data points, both PEDE and GOHDE models effectively resolve the Hubble tension, whereas $\Lambda$CDM predicts an intermediate $H_0$ between Planck and SH0ES. This highlights that a model’s ability to alleviate the tension depends strongly on the specific background data sets used.

Similar trends appear when comparing different BAO Ly$\alpha$ estimates. For instance, measurements from the SDSS survey—based on Ly$\alpha$ auto-correlation and Ly$\alpha$-QSO cross-correlation—favour a higher $H_0$, while their counterparts from DESI DR1 and DR2 do not. This further underscores the dependence of results on the chosen data set, with PEDE and GOHDE performing better under SDSS-derived quantities than under DESI in resolving the Hubble tension.

A key advantage of the BAO Ly$\alpha$ data lies in the fiducial cosmology assumed when deriving the $H(z)$ values. The original analyses assume a matter density $\Omega_m = 0.27$, which enables PEDE and GOHDE to predict a higher $H_0$. This reveals that the inherent effects of the data are tied to correlations among free parameters of the model; a lower $\Omega_m$ naturally prefers a larger $H_0$, since the model-independent quantity constrained by observations is $\Omega_m H_0^2$.

At this point, we emphasize again that it is the effective matter density that plays the key role, since our focus is solely on background evolution. For the GOHDE model, although we estimate $\Omega_m$ from the data, the dynamics are actually governed by the effective matter density, $\tilde{\Omega}_m$. This distinction is particularly important when interpreting data points such as the CMB shift parameter, which requires information both with and without dark energy. In our analysis, we assume all models reduce to CDM behavior when dark energy is absent. For PEDE and $\Lambda$CDM, this means considering only the fluid components that scale like matter. However, for GOHDE -- where dark energy acts as a tracking solution -- during the matter-dominated era it behaves similarly to CDM. Therefore, instead of simply excluding dark energy, we rewrite GOHDE as an effective $w$CDM model and identify the matter-like scaling component to determine the CDM contribution. With the exception of contour plots, all the tables report the effective matter density $\tilde{\Omega}_m$ when discussing the GOHDE model.

This brings us to the impact of including the CMB shift parameter and Pantheon$^+$ supernova data in our analysis. Since neither the CMB shift parameter nor Pantheon$^+$ alone can reliably estimate $H_0$ without a proper prior, we combine them with cosmic chronometers (CC) and BAO data. This combined analysis reinforces our earlier conclusion about the importance of parameter correlations: the CMB shift parameter places strong constraints on the matter density, enabling PEDE and GOHDE models to perform well. Although Pantheon$^+$ can also constrain $\Omega_m$, it does so only up to the combination $\Omega_m H_0^2$, as $H_0$ and the absolute magnitude $M$ remain degenerate. Consequently, the value of $H_0$ remains largely undetermined when using Pantheon$^+$ alone. This leads to an order of magnitude difference in the precision of $\Omega_m$ estimates from Pantheon$^+$ compared to Planck 18. As a result, PEDE and GOHDE models -- which showed strong performance in alleviating the Hubble tension when BAO Ly$\alpha$ data were used -- perform significantly worse once Pantheon$^+$ data are added. However, when the CMB shift parameter is also included, the matter density becomes tightly constrained, and the models’ ability to address the Hubble tension becomes evident.Since $\Omega_m H_0^2$ is tightly constrained by various datasets, introducing an additional free parameter is necessary to break the anti-correlation between them. This parameter must achieve this decoupling without compromising the overall consistency of the model.

Exploring the model’s features reveals that, by default, it cannot resolve the Hubble tension. When the free parameter is not fixed to any specific values, the data tends to prefer values that yield a lower Hubble parameter, which diminishes the model’s effectiveness. The models only demonstrate their potential to alleviate the tension when the free parameter is fixed at certain specific values. While this need for fine-tuning can be seen as a disadvantage in principle, there is a positive aspect if these specific values can be interpreted as meaningful phenomenological choices rather than arbitrary adjustments.

We further examined the models within the interaction framework. Upon reparameterization, we identified specific traits in the parameter space that enable the models to address the Hubble tension. One notable feature is the divergence in interaction strength and a reversal in the energy flow direction between components. While this divergence suggests infinite interaction -- an unphysical concept -- our exploration of horizon entropy provides a different perspective.

We observed that this flip coincides with a violation of the second law of horizon thermodynamics. This onset of violation manifests as a sign flip and divergence in the interaction strength, offering a novel interpretation not previously reported in the literature, to the best of our knowledge. These phenomena are inherently tied to the phantom-like behaviour of dark energy. Importantly, this analysis indicates that it is not the past or future phantom nature of dark energy that distinguishes these models, but rather the specific onset of the second law violation associated with the Hubble horizon.

From this, we conclude that a future violation of the second law is an inevitable feature of models capable of resolving the Hubble tension. This can also be interpreted as a turning point in the Hubble flow. Unlike the $\Lambda$CDM model, which does not exhibit such turning points, these models predict a future increase in the Hubble parameter after a certain period. Observations of past turning points would automatically disqualify the models discussed here. 

The concept of turning points in the Hubble flow has been previously discussed in the context of conventional holographic dark energy models, such as in \cite{Colgain2021critique}. While the GOHDE model has its foundations in holographic approaches, our findings predict a turning point in the future. Consequently, the violation of the null energy condition -- another critical feature -- is deferred to the future in the models analysed here.

In the course of concluding our study, we encountered several related works that also emphasise the importance of these data points under different contexts \cite{akarsu2025dynamical,sharov2018predictions}. While our findings regarding the impact of BAO Ly$\alpha$ data points align with these studies, \cite{sharov2018predictions} reports a reduced influence on the Hubble parameter. This discrepancy arises from their estimation of a significantly high negative curvature, where additional correlations play a crucial role in shaping the results.
Moreover, the most comprehensive approach would involve using the full distance priors, as suggested in \cite{Chen_2019}, or analysing the models with the complete CMB data to achieve a more thorough understanding. However, such an analysis lies beyond the scope of this manuscript, as our focus remains restricted to the background evolution. These aspects are reserved for future investigations.

\begin{acknowledgments}
I sincerely thank Titus K. Mathew and N. Shaji for their valuable guidance and support throughout this work. I am grateful to Rhine Kumar A. K. and the Nuclear Physics Group at CUSAT for providing access to their workstation, which was essential for the analyses in this study. I deeply appreciate Sarath Nelleri for the insightful discussions on the PEDE model. My thanks also go to Jackson Levi Said, Rafael C. Nunes, Eoin Ó Colgáin, and Sunny Vagnozzi for their helpful suggestions and for pointing out key references. A special thanks to Rahul Shah for identifying typos in the equations in an earlier version of the manuscript. \\
\end{acknowledgments}

\paragraph*{\textbf{Use of AI tools:}} I acknowledge the assistance of ChatGPT 3.5 and Grammarly in improving the grammar and English language usage in this article. After utilising these tools, I thoroughly reviewed and edited the entire content, taking full responsibility for its accuracy and quality.

\paragraph*{\textbf{Python modules used:}} Numpy \cite{harris2020array}, Pandas \cite{reback2020pandas}, Scipy \cite{2020SciPy-NMeth}, Emcee \cite{Foreman-Mackey_2013}, Lmfit \cite{mattnewville20238145703}, PyMC \cite{abril2023pymc}, Scienceplots \cite{SciencePlots}, Corner \cite{corner}, Seaborn \cite{Waskom2021}, and Matplotlib \cite{Hunter:2007}. The datasets used in this study were identified in advance, and the MCMC algorithms were executed on a moderately equipped workstation with 12 cores and 64 GB of RAM. The codes and datasets employed in this analysis are available upon reasonable request.

\paragraph*{\textbf{Data Availability Statement:}}
This manuscript has no associated data, or the data will not be deposited. [Authors’ comment: The datasets used are cited within the manuscript, and all MCMC chains generated can be made available upon request via email to the author.]

\paragraph*{\textbf{Code Availability Statement:}}
This manuscript has no associated code or software. [Authors’ comment: The codes employed in this analysis adhere to standard Python protocols for MCMC analysis, as referenced in the section “Python modules used.” There are no additional codes or software to disclose publicly. Specific codes and details of the analysis can be provided upon reasonable request. Please direct such requests to the author via email.]

%

\end{document}